\documentclass[%
 aps,
 showkeys,
 amsmath,amssymb,
 superscriptaddress, 
 floatfix,
]{revtex4-1}

\usepackage{graphicx}
\usepackage[caption=false]{subfig}

\usepackage{lipsum}
\usepackage{units}

\usepackage{hyperref} 
\hypersetup{
    colorlinks=true,
    linkcolor=blue,
    filecolor=magenta,      
    urlcolor=black,
    citecolor=black,
    pdftitle={Control Systems and Data Management for High-Power Laser Facilities},
    bookmarks=true,
}
\usepackage[open, openlevel=2]{bookmark}
\usepackage{setspace}

\usepackage[capitalize]{cleveref}

\begin{document}

\title{Control Systems and Data Management for High-Power Laser Facilities}

\author{Scott Feister}
\affiliation{California State University Channel Islands, Camarillo, California 93012, USA}
\email{scott.feister@csuci.edu}

\author{Kevin Cassou}
\affiliation{Université Paris-Saclay, CNRS/IN2P3, IJCLab, 91405 Orsay, France}

\author{Stephen Dann}
\affiliation{Central Laser Facility, STFC Rutherford Appleton Laboratory, Didcot OX11 0QX, United Kingdom}

\author{Andreas D\"{o}pp}
\affiliation{Ludwig--Maximilians--Universit{\"a}t M{\"u}nchen, Am Coulombwall 1, 85748 Garching, Germany}

\author{Philippe Gauron}
\affiliation{Université Paris-Saclay, CNRS/IN2P3, IJCLab, 91405 Orsay, France}

\author{Anthony J. Gonsalves}
\affiliation{BELLA Center, Lawrence Berkeley National Laboratory, Berkeley, California 94720, USA}

\author{Archis Joglekar}
\affiliation{Department of Nuclear Engineering and Radiological Sciences, University of Michigan, Ann Arbor, Michigan 48109, USA}

\author{Victoria Marshall}
\affiliation{Central Laser Facility, STFC Rutherford Appleton Laboratory, Didcot OX11 0QX, United Kingdom}

\author{Olivier Neveu}
\affiliation{Université Paris-Saclay, CNRS/IN2P3, IJCLab, 91405 Orsay, France}

\author{Hans-Peter Schlenvoigt}
\affiliation{Helmholtz-Zentrum Dresden -- Rossendorf, Bautzner Landstraße 400, 01328 Dresden, Germany} 

\author{Matthew J. V. Streeter}
\affiliation{Queen’s University Belfast, BT7 1NN, Belfast UK}

\author{Charlotte A. J. Palmer}
\affiliation{Queen’s University Belfast, BT7 1NN, Belfast UK}

\date{\today}

\begin{abstract}
The next generation of high-power lasers enables repetition of experiments at orders of magnitude higher frequency than was possible using the prior generation. Facilities requiring human intervention between laser repetitions need to adapt in order to keep pace with the new laser technology. A distributed networked control system can enable laboratory-wide automation and feedback control loops. These higher-repetition-rate experiments will create enormous quantities of data. A consistent approach to managing data can increase data accessibility, reduce repetitive data-software development, and mitigate poorly organized metadata. An opportunity arises to share knowledge of improvements to control and data infrastructure currently being undertaken. We compare platforms and approaches to state-of-the-art control systems and data management at high-power laser facilities, and we illustrate these topics with case studies from our community.

\end{abstract}


\maketitle

\begingroup
\setstretch{1.25}
\tableofcontents
\endgroup

\section{Introduction}

\subsection{Shifts in High-Power Laser Technology Necessitate Revised Digital Infrastructure}
High-power and high-intensity laser-plasma interactions provide a versatile experimental platform. They can produce extreme plasma environments, either for laboratory astrophysics and fundamental plasma physics, or as a unique source of secondary radiation.  Secondary sources include bright, keV - MeV X-rays \cite{Corde2013RMP}, low-emittance and high-current electron beams \cite{Couperus2017NC,Gotzfried2020PRX,Foerster2022PRX}, GeV electron beams \cite{Leemans2014PRL,Gonsalves2019PRL,Kneip2009PRL,Kim2021AS}, ultra-short MeV proton beams \cite{Badziak2018JP, Schreiber2016RSI} and pulsed neutron sources \cite{Alejo2015}. These sources have demonstrated significant potential for applications \cite{Albert2014PPCF} including rapid, high spatial resolution x-ray tomography \cite{Wenz2015NC,Cole2015SR,Dopp2018Optica}, free-electron lasing \cite{Wang2021Nature,Pompili2022Nature,Graydon2022NP}, FLASH radiotherapy \cite{Esplen2020, Chaudhary2021}, and materials damage testing \cite{Zhang2018OE}.  
In order to develop these sources for applications (e.g. optimizing their stability and tunability), and in order for them to be competitive with alternative sources, it is necessary for the source repetition rate to drastically increase from sub-Hz to hundreds-of-Hz (and beyond).

Tackling the obstacles to achieving multi-Hz~repetition-rate high-intensity laser-plasma interactions has been a focus of the high-power laser community in recent years. Great progress has been shown in laser technology \cite{Zhaoyang2022}, replenishing targets \cite{Prencipe2017, Chagovets2021, George2019, Treffert2022, Kraft2018, Puyuelo2019,Oertel2019} and online diagnostics \cite{Grace2021PPCF, Obst2018PPCF, Waxer2018, Downer2018RMP}.  The increasing availability of experimental facilities compatible with high-repetition-rates now highlights the need to adjust traditional experimental practices, and control, in order to fully exploit the opportunities offered by these systems \cite{Hatfield2021NAT}.  Among these needed adjustments, new systems should enable \textit{automation} of scans of experimental and laser parameters, rather than relying on repeated re-configuration through manual user input.  In addition, data from a large suite of diagnostics should be acquired and collected at least at the repetition rate of the slowest experimental component (e.g. laser, target), ideally with the associated metadata to enable efficient online analysis.  Such systems can then be used for automated communication between diagnostic elements and control elements, enabling control without a human ``in-the-loop'' \cite{Dopp2022arxiv}.  In recent years this has been demonstrated in proof-of-principle experiments \cite{Loughran2023HPLSE, Jalas2021PRL,Shalloo2020NAT,Dann2019PRAB} illustrating the deeper insight and source enhancement that is enabled by such a shift in methodology.  

With a new generation of high-repetition-rate, high-power lasers nearing completion, re-tooling is underway in many areas linked to high-power laser facilities and their experiments. These areas include targetry, scientific instruments, data pipelines, mechatronics systems, analysis software, data pipelines, and data management. Recently, two special issues of peer-reviewed journals have been dedicated to ``Target Fabrication'' \cite{Spindloe2018preface} and ``The High Repetition Rate Frontier in High-Energy-Density Physics'' \cite{Heuer2022preface}. This manuscript addresses a related topic -- facility control systems and data management -- which is the digital infrastructure upon which scientific experiments are designed and executed. 

\subsection{Approaches to Tooling}

\begin{figure*}[htpb!]
\begin{centering}
\includegraphics[width = 17 cm]{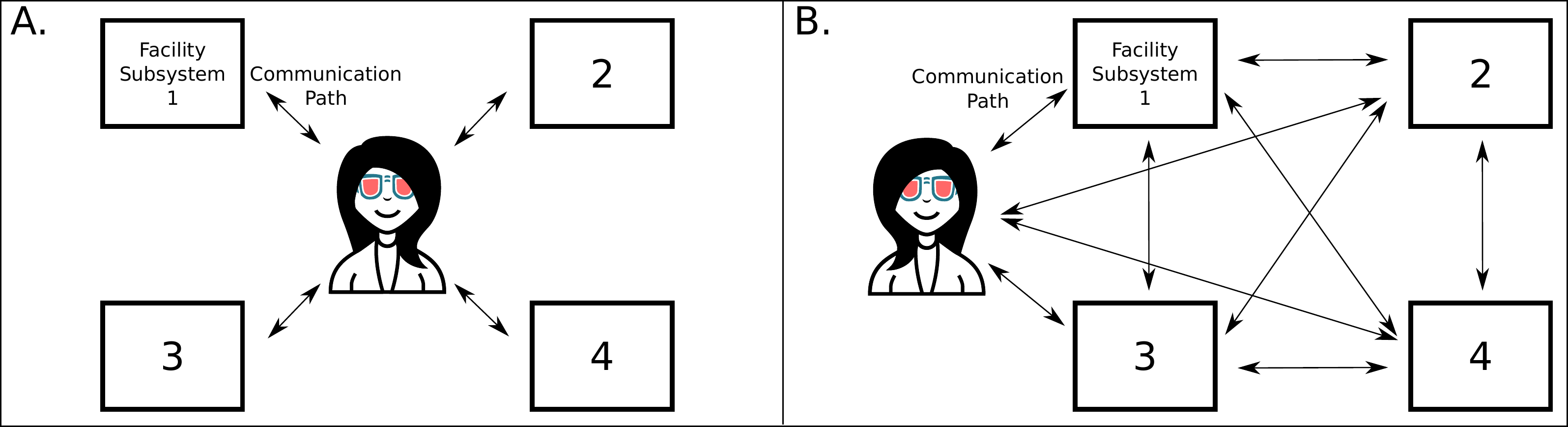}
\caption{A) Diagram of a high-power laser facility without a laboratory-wide control system. People tie the subsystems together. For example, a person might adjust parameters on the laser amplifier subsystem in response to  observations from the target-chamber subsystem. B) Diagram of a facility with a laboratory-wide control system. Consistent implementation throughout the facility opens new communication pathways between subsystems. These new pathways enable laboratory-wide automation and control feedback loops without human mediation. Humans retain communication with all subsystems.}
\label{fig:humancontrol}
\end{centering}
\end{figure*}

An effective laboratory control system can enable facility-wide communication between instruments, sensors, software, and humans. To clarify each category in the context of high-power laser facilities: (1) Instruments at high-power laser facilities could include, \textit{e.g.}, laser power meters, laser contrast diagnostics, particle time-of-flight detectors, and electron spectrometers. (2) Sensors could include, \textit{e.g.}, thermometers and pressure sensors. (3) Software could serve roles of, \textit{e.g.}, laboratory automation, data storage, human-interfacing, and online data analysis. (4) Humans involved could include, \textit{e.g.}, facility staff, principal investigators, graduate students, postdoctoral scholars, and visiting scientists.

One common approach to control in a high-power laser facility is to leave various elements semi-isolated from one another, with humans as the direct mediators of communication between various facility elements. Elements may be computer-interfaced, \textit{e.g.} through vendor-provided graphical software or through a custom LabVIEW interface, but not in digital communication with one another. In this approach, a person is responsible for synthesizing data inputs from various elements, and making the appropriate system re-configurations between repetitions (i.e. laser shots). (See Fig. \ref{fig:humancontrol}a for an illustration.) The increase in repetition rate of high-power laser facilities necessitates a need to move beyond this human interfacing, towards digital interfacing. This leads to adoption of control methods that are managed primarily in network-interfaced software (which itself is configured by humans). (See Fig. \ref{fig:humancontrol}b.) A distributed networked control system\cite{Ge2017IS} is one in which the laboratory elements interact between one another (rather than to one central person or even one central control hub), and it is designed to be fault-tolerant and scalable. It may implement strategies (such as redundancy) to ensure communication and control can continue even when confronted by network disruptions or other issues.

Data management is another area of digital infrastructure in a laboratory where the approach may be organized, disjointed, or somewhere in between. One common approach to data management in a high-power laser facility is to maximize flexibility by utilizing humans at all stages of data management. For example, team members might initiate data collection on many computers separately and keep track of the relationships between that data in a laboratory notebook. Metadata -- or the data which provides context for the data itself -- could include information such as environmental parameters, laboratory configurations during data collection, and the experimental diagnostics settings. This metadata is manually captured and recorded by a human operator, then collected together with the data and labeled in such a way that it can be re-associated with the relevant data. In this approach, when other scientists wish to leverage that data in analysis, people communicate the data and metadata to collaborators via email, shared cloud folders or other means.

This approach to data management for high-power laser facilities has relied on sufficient and accurate human-described context associated with each element of data. However, when lasers are pulsed/fired at many times per hour, less time is available to capture the metadata -- and thus, there is more opportunity for mislabeled or missing data. Human error is also an issue in low-repetition-rate experiments, and so automated recording of metadata can benefit even these less-frequent measurements. However, in the most extreme cases of new high-power laser systems capable of generating terabytes to petabytes of data in a span of days or weeks, it is not even feasible to copy the raw data onto a personal computer, let alone easily distribute it in-full to colleagues. Isolated, fragmented copies of partial datasets could be expected to result from person-to-person distribution. We suggest that organized approaches to data management for the next generation of high-power laser facilities involve people stepping back from direct manipulation and distribution of data. So, we propose approaches to data management with systems that: 
(1) reliably save data with unambiguously associated metadata; (2) provide easy access to the data from multiple users and institutions; (3) include modular software tools for analysing data.

Many laboratories in our community have already implemented facility-level approaches to control systems and data management. We share several examples as case studies throughout this article, and hope that the community can seize this opportunity to learn collaboratively and maximize the utility of the next generation of digital infrastructure.

\subsection{Case Study: Draco laser at HZDR} \label{scn:hzdr}
This community example adds concreteness to the challenges of upgrading existing digital infrastructure, which may be disparately organized.

\paragraph{How did multiple subsystems develop at this facility?} 
Helmholtz-Zentrum Dresden--Rossendorf (HZDR) was originally founded as a nuclear physics research center in mid-1950s and hosts nowadays, together with several other facilities, the Draco laser system \cite{Schramm2017}, in operation since 2007, and the PEnELOPE system \cite{Albach2019, Siebold2013}, in commissioning. Research into relativistic plasmas was initiated at HZDR in the mid-2000s. Laser systems were installed for synergy reasons as additions to the existing accelerator-driven photon and particle source called ELBE \cite{Michel2016}, sharing structural shielding, building infrastructure as well as the radiation monitoring and interlock system of ELBE. Further control systems are complete additions due to the higher flexibility of a laser-plasma  experimental arrangement compared to a fixed accelerator machine. These systems were either newly developed or were part of the corresponding subsystem itself (e.g. laser control system by Amplitude and Scarell). The newly developed components were conceived as stand-alone systems for swift start of operations. Due to the external development of the laser control system, an approach in a single framework was not possible or desired. Instead, subsystems were developed step-by-step in a historical progression, partly in parallel, and partly by vendor.

Subsystems were each developed to suit different needs. For example, the vacuum control system in a laser-plasma experiment can have less strict interlock requirements than for a superconducting accelerator like ELBE, but it must allow for frequent pump-down cycles with a simple user interface. 
Likewise, in contrast to the more long-term fixed setups found at a typical accelerator facility, actuators in laser-plasma experiments are frequently re-configured and assembled with new components.

Despite systems being based on different frameworks, subsystems are partly interfaced to one another, and re-usability and uniformity within a subsystem, where possible, is implemented. As an example thereof, the laser-plasma data acquisition is often done with cameras. The software Laser Light Inspector \cite{llinspector} provides a common user interface for various camera vendors with a number of live analysis features, and was therefore chosen by HZDR for camera acquisition control. It further provides a centralized interface for remote control of the camera clients. This reduces efforts for e.g. file path configuration etc. significantly since a vast majority of detectors is operated via this software, and leads to a high level of homogeneity within this subsystem.

Likewise, acquired images are always stored locally at each client, like the native vendor software does for other detectors like optical spectrometers or oscilloscopes, not being integrated in Laser Light Inspector. As result, dozens of acquisition PCs are deployed, each operating a few cameras and/or other detectors and collecting data locally for all detector types. Although imperfect, this methodology does offer high flexibility and redundancy, since virtually any camera or device can be connected to any nearby PC. Locally-stored data files are regularly copied over the network to a central data repository by a file synchronization software \cite{ffsync}.

\paragraph{How do humans execute and process each shot?}
The above operation mode, involving dozens of dedicated acquisition computers, has led to quite efficient subsystems with a high degree of flexibility, but requires significant human interaction and consequently imposes a high work load on the scientists. There is no assigned staff operator crew to tie together the local subsystems as one would find at the larger accelerator facility.

\begin{figure*}[htpb!]
\begin{centering}

\subfloat[]{\includegraphics[width = 17 cm]{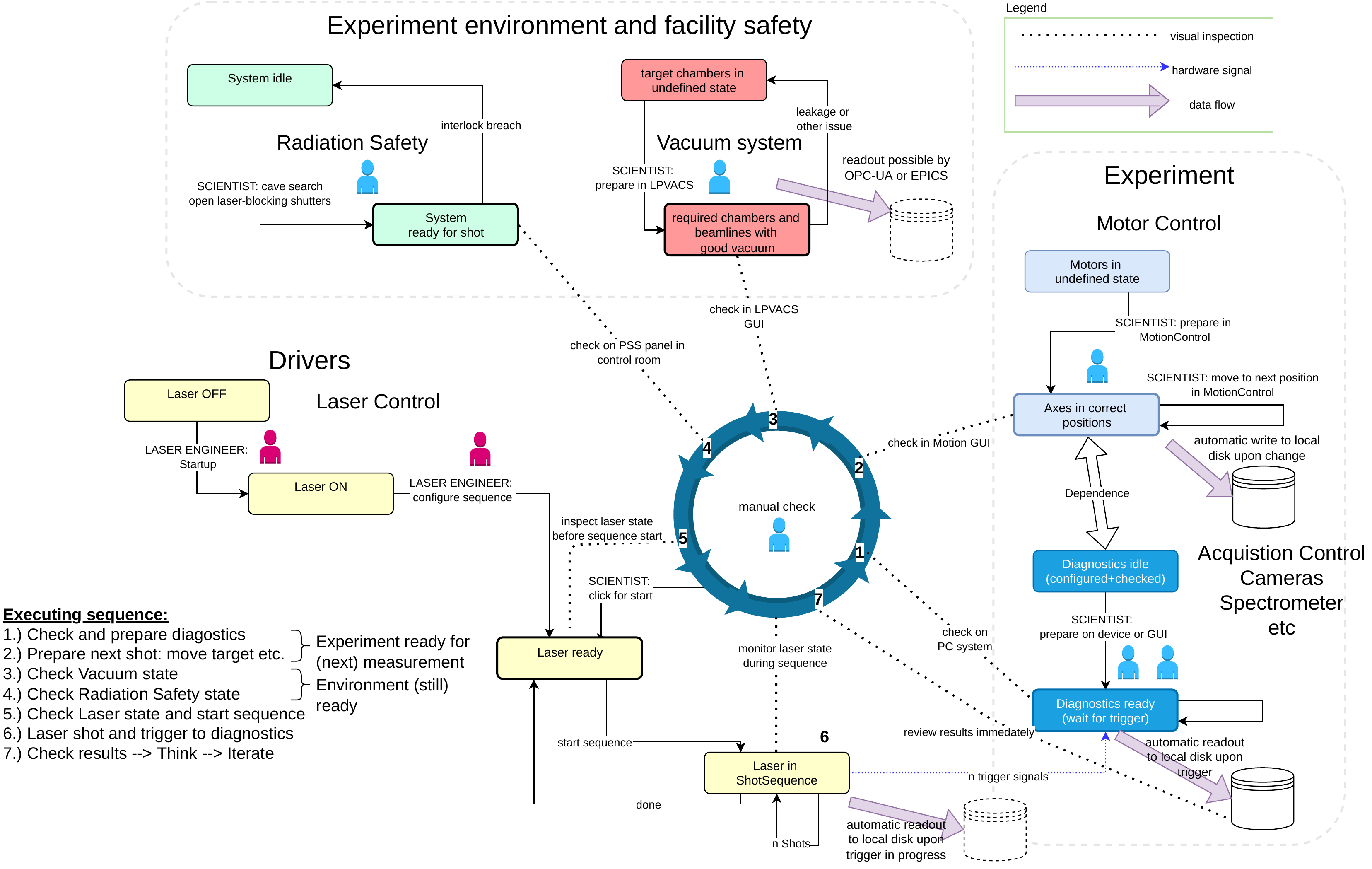} \centering \label{fig:hzdrmain}}

\subfloat[]{\includegraphics[width=6.5 cm]{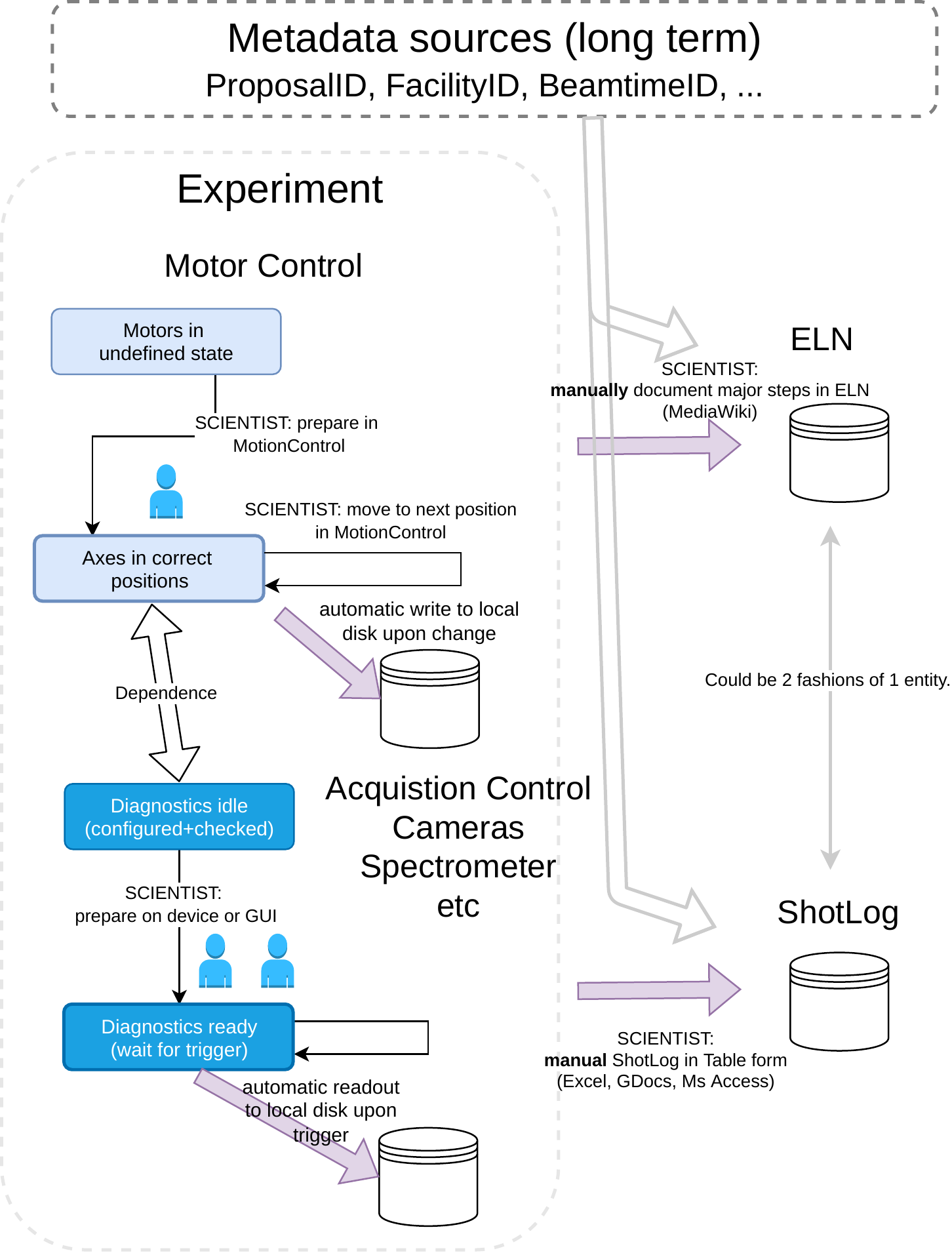} \centering \label{fig:hzdrmeta}}

\caption{Engineers and scientists (represented with maroon and blue icons) manually conduct steps in the operational sequences for Draco at HZDR. A) (Top) Facility laser shot sequence diagram for ion acceleration experiments. Note the many steps in which people tie together subsystems. The central blue circle indicates the laser shot sequence, executed and surveilled by scientist. Automated storage is part of many subsystems. B) (Bottom) Manual logging to complement and complete the data storage, to curate and enrich with metadata.}
\label{fig:hzdr}
\end{centering}
\end{figure*}

Fig. \ref{fig:hzdr}a illustrates the flow of operations for laser-driven ion acceleration experiments, typically done at a shot-on-demand level, i.e. about 1 shot per minute and parameters are deduced from previous shots. Laser-driven electron acceleration experiments operate similar but typically perform a series of shots at one point in parameter space with repetition rates between \unit[0.1-0.5]{Hz}, before changing the parameters.

In the shot preparation phase, the numerous subsystems are prepared by staff and scientists, partly in parallel. This includes evacuation of experimental chambers, initializing the detectors, searching and (inter)locking of the experimental rooms, and preparation of laser to operate at high pulse energy.

The actual high-power laser shots are performed as a sequence of steps (shown as a large circle in the center of Fig. \ref{fig:hzdr}a and enumerated 1-7), coordinated by the scientists, which partly transfer status information between the subsystems but also process acquired information and plan the course of action. At first, diagnostics need to be checked and armed (e.g. save on trigger, correct destination file path etc) and (2) the experiment must be in the correct positional state (e.g. correct target and detector settings). Then all facility and safety interlocks (3 and 4) must be fulfilled. If everything indicates a ``go'', a click in the laser control software starts a shot sequence (5).

The laser shot sequence consists essentially of a well-timed series of trigger signals (6), controlling processes within the laser system (e.g. flash lamps and Pockels cells) to produce the energetic laser pulse as well as triggering all detectors at the experiment with the appropriate delay such that they can record the ultrafast interaction.
That trigger signal from laser to experiment (blue dotted line in Fig. \ref{fig:hzdr}a) is so far the only direct and automatic information transfer between subsystems, not mediated by scientists.

After that sequence, the immediate results can be viewed (7) and the cycle can start again with a new target and refined detector parameters. This step is not so much information transfer between the systems like the other steps, but rather information processing and decisions making, where scientist are mandatory.

As shown in Fig. \ref{fig:hzdr}a, subsystems do have automatic saving or logging, but since they are all autonomous, each subsystem's storage is independent of each other and they therefore need some relation to each other. In addition, not everything can be automatically logged, e.g. observations and reasonings. Hence there is the need for a second layer of that schematics for manual logging and tying relationships. That is depicted in Fig.\ref{fig:hzdr}b which complements the right part of Fig.\ref{fig:hzdr}a, where the primary experimental data storing is shown; data from the experimental environment like the vacuum system is rather secondary in this regard.  Fig. \ref{fig:hzdr}b shows in particular the importance of manual logging, i.e. documenting all work of setup and calibration into an electronic lab notebook (ELN), as well as documenting the shots into a shot log. The shot log is very important for later analysis of experiments as it documents the key experimental parameters and observations, for example diagnostic filter settings or unexpected laser behaviour. It also provides, via shot counts and timestamps, the relations between the automated logs. Finally, the connection to further metadata, including descriptors for the entire experimental run, happens at that manual stage. However, those relations are to a large extent only visible for humans.

The ELN, realized by a Mediawiki system \cite{MediaWiki}, allows for the full and rich documentation of the facility, in particular set-up, development, maintenance and calibration works, but can also store or link the experimental shot data.

\paragraph{Are there plans to move toward a facility-level control system?}
HZDR's laser experiments deploy a number of different control systems, based on various software frameworks, partly for historic reasons. The ansatz of HZDR is to arrange the subsystems in a coherent way, rather than re-building the system in a single framework. Coupling is currently established via scientists and interfaces, and will always be for sake of flexibility. The goal for developments is rather to add interfaces to the subsystems where necessary and to add agents in between, as shown in Fig.\ref{fig:humancontrol}, such that more and more automation can take place where the established workflows allow for.  HZDR respects the fact that there will always be some new development that should be quickly embedded into experiments, hence the systems should have interfaces at very general, low-level tiers, such that scientists can realize the changes of the control system. Following are a few examples of existing plans by this facility to reduce reliance on human interfacing between subsystems and increase coherence:

\begin{itemize}
    \item To reduce the work load of status monitoring during the shot cycle, a handler could generate a joint signal from the vacuum control system and radiation safety system, that could be fed to the laser system to inhibit the shot sequence if either subsystem is not in the proper state.
    \item The laser system could generate a common shot ID that can be fed to the diagnostics. It might be necessary to add an agent which could add that ID to all file names of files generated from diagnostics. This would keep the primary acquisition routine as is and add valuable metadata to all kinds of detection.
    \item Upon synchronizing the files to a central repository, they could be parsed to extract metadata which is currently encoded in the file path and name. That metadata could be sent to a database like SciCat \cite{scicat}, and that database could be joined with that of the shot log. Again, such approach would be independent of the acquisition software and therefore very general.
    \item On individual diagnostic computers can specialized analysis scripts be developed and deployed online during the experiments. Analyzed data can be forwarded via messaging protocols to a flexible database such as MongoDB\cite{MongoDB} and visualized via Grafana\cite{Grafana} during experiments. This can help to literally better see changes and relations of otherwise tabular information.
\end{itemize}

These adaptations are a step in the right direction towards reducing human interfacing. They will help streamline open-loop operations at moderate repetition-rates. However, without a move to a distributed networked control system, such a system is not ready for facility-level closed feedback loops. And, although metadata has been considered, this system relies on files passed across a network, and is not configured with high-speed data analysis pipelines for high data rates.

\subsection{Digital Infrastructure for next-generation facilities}
The case study from HZDR shows that historically grown control systems may operate efficiently and with high flexibility at moderate repetition-rates. But they also become increasingly complex over time, requiring significant human resources for interfacing between individual parts. 
Higher data rates are a necessary step forward to provide deeper insight into high-power laser-plasma interactions and to produce competitive secondary radiation sources for applications. But moving toward increasing number of diagnostics and higher repetition rates pushes the boundaries of (too) heterogeneous approaches. 
Instead of being monolithic, many modern control systems are based on modular approaches, allowing them to scale up while keeping complexity at a reasonable level and allowing them to quickly adapt to changing requirements. The modularity also allows for different levels of communication, from low-level device access to high-level user interfaces.
However, the choice of an adequate, future-proof control system is also a lasting one. Transitioning to a new control system may take years and requires considerable effort. Many facilities in our community are faced right now with choices that will impact their operation for the next decades, as they develop new digital infrastructure to meet the needs of the next generation high-power laser experiments. In the next sections, we compare platforms and approaches to next-generation infrastructure for control systems and data management at high-power laser facilities, and illustrate these topics with case studies from our community.

\section{Approaches to Laboratory Control}

While there are clearly available strategies for developing control systems to make full use of the multi-Hz data acquisition rates, realising them can be a complex task in itself. Here (\cref{scn:control_opportunities}), we illustrate this by considering opportunities enabled by organized approaches to laboratory-wide control and several very real challenges inhibiting adoption of these new approaches within the high-power laser community (\cref{scn:control_challenges}).

\subsection{Opportunities}\label{scn:control_opportunities}

\subsubsection{Laboratory-Wide Automation}
When controls are operating in isolation, a human being is required to make adjustments to various systems within the laboratory. When controls are integrated into a digital control system that is organized throughout the laboratory, automation across sub-systems is possible.
For example, a laser intensity scan can be executed and may involve adjusting pump laser diode timings in the amplifier, translating a motorized stage in the compressor chamber, tilting a mirror in the target chamber, \textit{and} increasing the image gain setting on a scientific camera. 
Laboratory-wide automation is important for scientists because it enables execution of a planned shot sequence in which parameters are adjusted across multiple laboratory subsystems, \emph{without} pausing work to wait for humans to make manual adjustments. As a consequence automation of a shot sequence can dramatically speed up the overall scientific experiment and free up time for tasks requiring the specialist skills of a scientist. 

\subsubsection{Laboratory-Wide Feedback Control Loops}
A major opportunity of extensive automation is to ``close the loop'' across an entire experiment enabling automated control based on scientific data. Data acquired from measurement of laser-plasma interactions can be piped into the control system to make control decisions. 
Importantly, with a control system implemented in a organized way, scientists have wide flexibility to decide which controls will be manipulated, and a common interface with which to make changes to the control settings. 
With a human out of the control loop, feedback-driven decisions can be made at much higher repetition rates and with reduced human bias.

\subsection{Challenges}\label{scn:control_challenges}
\subsubsection{Modularity and Network Bandwidth Management}
With abstractions in place between device hardware and control interfaces, one can more easily swap out low-level hardware without breaking the high-level control interfaces. For example, one might upgrade camera hardware and switch to a different device driver, without impacting the software for display interfaces or image analysis. This leads to increased modularity of devices in the laboratory, since software for display, storage, and analysis can be isolated from the complexities of device hardware.

When software for display, storage, and analysis are separated from one another and communicating via common protocols, this allows for modularity in the software. For example, multiple different analysis pathways can be built which expose the same public interface, such that they are interchangeable in the experimental data pipeline, to allow comparison and confidence in each analysis. This provides flexibility, allows for re-use of software components (e.g. data storage modules can be re-used with small adaptations in similar subsystems), and can in some ways bring more order to the system as there are fewer hidden, unknown connections between software components than one might find in a monolithic codebase. However, it also raises the challenges of organizing various versions of these modules, and keeping track of changes to public interfaces and how this affects dependent modules, since the software modules are not kept globally in lockstep. Fortunately, this challenge is common in non-scientific software development, and there are software development best practices for managing dependencies and establishing versioning to help manage the added complexity of interacting modules. Separation of data display, storage, and analysis across a local network can result in high network data rates, especially if large images are being streamed between multiple points in a local network. One approach to mitigate this issue without losing the benefits of modularity is to build modular software components, communicating via public interfaces, but have them running on the same computer (or a small local subnetwork) so that laboratory-wide network bandwidth is not an issue. A second approach is to send reduced data across the network; for example, the polynomial fitting coefficients to a curve. A third approach is to be selective about which data is rejected and which data is transferred or stored. Approaches to down-selecting data within the data pipeline, including coordination of flagging certain data as noteworthy across the entire laboratory, are utilized at many large, data-heavy facilities. For example, the LHC ATLAS experiment involves “trigger chains” to reduce data streams through multiple levels of decimation, starting with hardware selection and ending with customizable software selection algorithms \cite{Jenni2003ATLAS, Berger2008ATLAS, Collab2020ATLAS}. As a second example, the LCLS-II Data System \cite{Thayer2017LCLSII} takes different approaches to configurable in-line data reduction. Though the technology developed for these experimental facilities may be oversized for many in the high power laser community, the ATLAS and LCLS-II examples can provide inspiration for smaller-scale event-selection techniques for data rate reduction, and provide evidence that managing network bandwidth is possible even with massive quantities of high-repetition-rate data.

At the level of challenges currently faced within the high power laser community, cameras acquiring digital images at high repetition rate present a unique challenge to a networked control system, and require special care during control system design to avoid overloading the network with image-data traffic. For example, a single monochrome, 8-bit, 10-megapixel image may represent 10~megabits ($\sim$1.3~megabytes) of data, and streaming images from two of these cameras with no image compression at a repetition rate of 50~fps would consume the entire theoretical bandwidth of a 1 Gbps network. One approach to managing cameras (without sacrificing image storage through event-selection) is to store images locally, and only send limited metadata about the stored image across the network (rather than sending the full images). An elegant implementation is found in the FACET-II system of SLAC \cite{Gessner2022facetii}, in which camera data is acquired by a control computer and then saved to network-attached-storage that is proximate to the control computer (the transfer occurs across only a single network switch). In this way, all heavy image traffic stays localized to the most-local network switch, and the overall network (which includes many distributed network switches) is not burdened. Similar to this approach, many single-board computers (such as the Raspberry Pi 4, NVIDIA Jetson, or Beaglebone AI-64) are now powerful enough to provide sufficient data processing for a single camera image stream, with on-board GPU analysis of images, and to host large storage devices like solid-state-hard-drives or external USB hard drives. These single-board computers can be configured to act as control computer, analysis computer, and network-attached storage, such that image data need not stream across the network at all, not even to a local network switch.

If common abstractions are adopted across multiple laboratories in the high-power laser community, this enables yet another opportunity to share devices across laboratories with minimal modifications to the digital infrastructure. This could facilitate travel to facilities with scientific instruments that plug into the digital infrastructure at that facility and work as they did at other facilities.

\subsubsection{Learning Curves and Inflexibility}
Unfamiliarity can impede the adoption of newer control systems, as stakeholders continue to ``use what they know''. Systems with robust software engineering foundations can require more advanced technical knowledge to develop, modify, and manage.  Steep learning curves can reduce flexibility, as stakeholders feel disempowered to modify their own control system. New control systems can be especially frustrating to team members who know how to build individual controls with their own tools of preference, but do not know how to integrate these into the control system. If these challenges are unaddressed, a high-quality control system architecture can inadvertently ossify the facility, serving as a barrier to entry for new devices and new team contributions. 

An idea for mitigating these challenges involving the ``human factor'' is to follow human-centered design \cite{Rouse2007book}.
In human-centered design, stakeholders are incorporated into the design process. Here, we provide one usable example of how human-centered design can be implemented to improve control systems in a high power laser laboratory. At a facility, a single person who is knowledgeable of the control system software (whether that be a dedicated software engineer, or a software-savvy graduate student) could block off one day per month in their schedule for observations. That software person could spend this one day per month observing graduate students, scientists, and technicians as they perform regular laboratory tasks. To ensure operations are not disrupted, the software person would take on an important responsibility to stay “out of the way” of stakeholders, such that the other members of the facility could continue using the facility as usual. Through observation, the software person could identify core tasks that various stakeholders need to do in the lab, which involve human-computer interaction, and keep that list up-to-date with each subsequent monthly observation. This document could be shared and asynchronously edited by stakeholders, or entirely ignored by stakeholders, depending on their preferences and schedules. The software person could furthermore document findings and adjust future versions of the software to fold in these observations. This repeated cycle of observation and improvement could result in overall improvement to the control system over the course of many months or years.

An example of a practical avenue for asynchronous communication of stakeholder needs would be building a single-click issue reporting mechanism in the control software (with this reporting system itself observed and streamlined to reduce frustration), such that issues can be created and then followed up on by the software person. Furthermore, the software person can regularly have short, informal conversations about experiences using the control system with stakeholders, and personally take on the burden of documenting these issues and experiences (rather than putting that burden on the stakeholders). Though implementing these ideas will require some time in the software person’s schedule, this time will be paid back by many small improvements to daily operations for the entire team at the facility. If human-centered design is implemented at multiple facilities in our community, sharing notes at conferences can help identify issues they may not have noticed, and sharing the methods tried to solve stakeholder problems can help the community converge towards common solutions to common problems.

Science at some laboratories requires more flexibility in the control system than others, and building the same degree of system inertia into control systems at two laboratories with different science cultures will lead to different results. Trade-offs between control system rigor and ease of prototyping can be considered for different scientific facilities, bringing the conversation to the forefront with stakeholders.
Human-centered design at all stages of development will help next-generation control systems meet the disparate needs of stakeholders in high-power lasers, including when compromises are required between stakeholder needs. At the facility level, as large systems are being designed, involving as many stakeholders as possible in the design stages of a control system allows for specifying better engineering requirements, which makes the control system better suited to all stakeholders.

\subsubsection{Timing and Synchronization}
Avoiding data synchronization errors at high-repetition-rate (including ``off-by-one'' data misalignment errors) requires several elements to work together flawlessly. First, each data element's timestamp must be locally precise to a significantly greater tolerance than the time between consecutive laser shots.  Second, each data element's timestamp must be globally accurate (across the entire laboratory), to similar minimum tolerances. Third, any internal delays for scientific instruments must be accounted for – as an example, consider a camera stream which feeds image buffers that are off-by-one, or instruments that return data at a longer delay than others.

Fortunately, timing and synchronization problems have been deeply considered and solved at many accelerator science facilities \cite{Hidvegi2011EuroXFEL, Gaget2022SARAF, icalepcs2019, icalepcs2021}. The high power laser community can learn from and adopt prior art in this area.

Two network timing synchronization protocols utilizing a facility’s Ethernet infrastructure are NTP (network time protocol), which aims to provide up to few-millisecond precision over a wide-area-network, and PTP (precision time protocol, IE1588) \cite{TI2012PTP, Krzyzanowski2021website}, which aims for sub-microsecond precision within a single local-area-network. For repetition rates up to about 1 Hz, time-stamping data using the control computer’s time (synchronized via NTP for sub-second accuracy) is sufficient. For rates up to $\sim$100 Hz, and for networks with consistent ping times less than 1~ms, time-stamping can still be feasible at the control-computer rather than within the data acquisition hardware, provided consistently fast processing of a consistent data stream from the data acquisition hardware. Specialized concurrent networking approaches and tools such as zeroMQ \cite{zeroMQwiki} can aid in synchronization.

For higher repetition rates, e.g. high power lasers operating at 1~kHz or 10~kHz, dedicated timing hardware is commercially available \cite{MRFpricelist, MicrosemiProducts}, and the data is typically time-stamped at the DAQ level (rather than relying on the control-computer operating system). An alternative solution to laboratory-wide time-stamping is to leverage available trigger counting available on many DAQ devices (or to write one's own trigger counting firmware for inexpensive microcontrollers), and to maintain a global laboratory trigger count rather than a global laboratory time.

\subsubsection{Legacy Instrumentation}

Existing equipment in laser-plasma facilities can present friction to the adoption of a laboratory-wide control systems.  This is particularly true of important legacy devices due to incompatibility with modern software or operating systems. However, this need not be the case.

Possibilities include finding existing device drivers or using vendor supplied software-development-kit which can be used to integrate the device properly into the control system. Alternatively, the legacy user interface (possibly a vendor-provided interface) can continue to be used with a patch between the legacy software and the laboratory-wide control system. For example, a commercial user interface for a camera may already permit saving images to a file, and one could write  an interface layer that reads the latest image from the file and makes it accessible on laboratory-wide control system. A last option would be to maintain legacy devices as they are, existing apart from the control system. The latter two approaches might be especially apt for devices requiring an old, insecure Windows computer without a network connection. 

\section{Comparison of Platforms for Laboratory Control}

While some facilities can offer significant research software engineering to support the development of control systems, many labs, in particular university-based systems, must adapt or develop their own systems.  Importantly, these must be manageable for researchers without a background in software engineering.

To date, a number of experiment control systems based either on expansion of previously existing experiment control systems at older laser facilities, or the adoption of control systems from beyond the laser community (e.g. High Energy Physics and photon science) have been explored. This section reviews the advantages and disadvantages of several of these, as well as the unique challenges facing the high-power laser-plasma community in control for future high-power laser facilities through specific examples.

\subsection{LabVIEW}
\subsubsection{Overview of LabVIEW}
National Instruments (NI) LabVIEW is a graphical programming environment, and is not itself a control system. However, it is already utilized in some aspect for controls at most high-power laser facilities in our community. LabVIEW is highly accessible to beginners and experts alike, and is widely used across many science domains. One major feature of LabVIEW, in the context of control system design, is the ease with which users can create interactive graphical interfaces. A second major feature is a robust library of instrument drivers. Many instruments used at high-power laser facilities already ship with vendor-provided LabVIEW drivers. A third major feature is a built-in infrastructure for shared variables across a network. As a drawback relative to open platforms, LabVIEW has licensing requirements which must be navigated for each deployed instance. Also, since it is a graphical programming language, code written in LabVIEW does not easily mesh with schemes for text-based version control.

Given the many LabVIEW users in our community, the reader is likely familiar with many features and downsides which have not been discussed in the preceding brief overview. A very large number of existing control systems in our community are based on LabVIEW. Some are built in a consistent fashion throughout the laboratory so-as to unlock all the high-data throughput, interconnected benefits of high-power laser facilities. The case study that follows illustrates how LabVIEW was incorporated into a home-grown control system architecture called GEECS at the BELLA center.

\subsubsection{Case Study: BELLA Center at LBNL}

\begin{figure*}[htpb!]
\begin{centering}
\includegraphics[width = 17 cm]{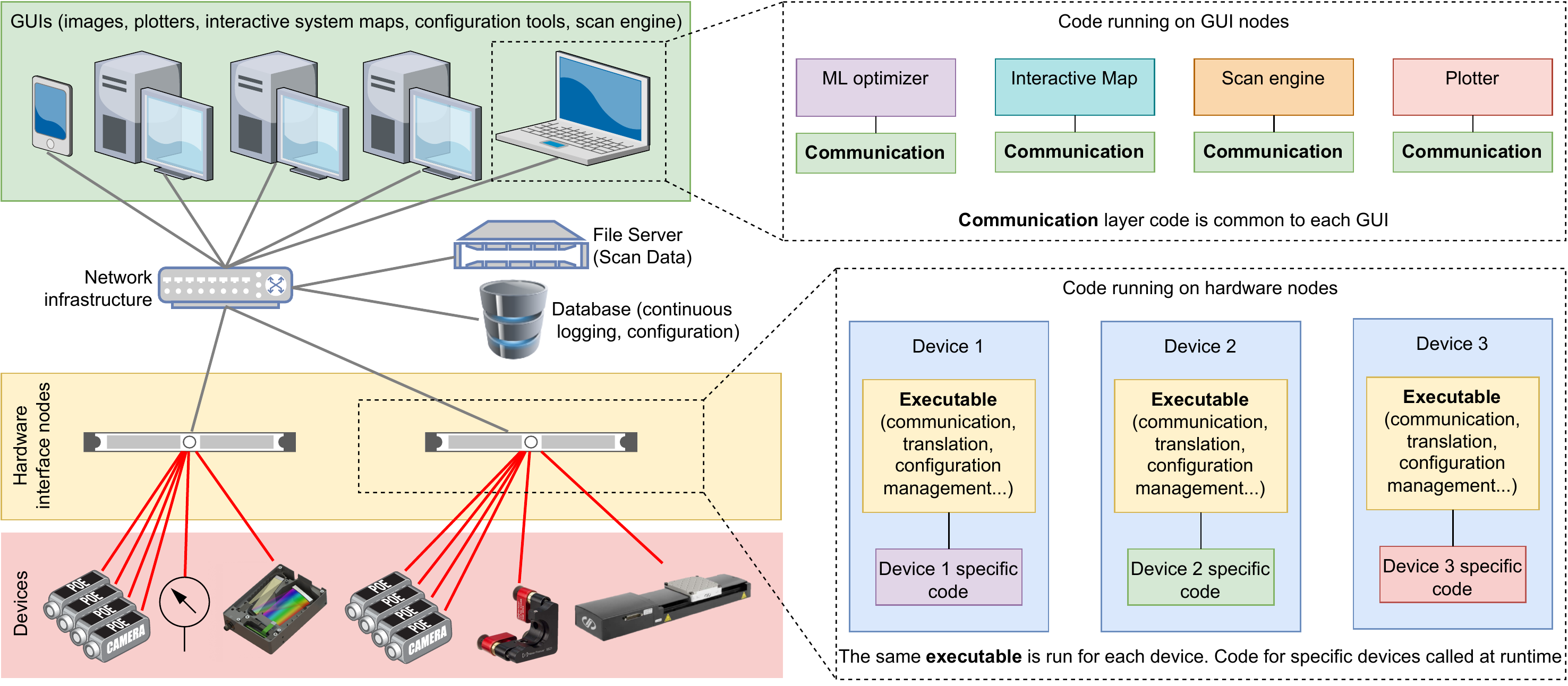}
\caption{Hardware and software overview of the GEECS control system, highlighting the layered modular approach. The hardware view is shown on the left, with multiple devices attached to a given computer, the number of which is determined by the system resources. The software running on the computers is shown on the right, which sends data and receives commands through the network infrastructure. The GUIs and the database receive data continuously, with the latter limited to scalar data. The file server receives scan data (which includes scalar, vector, and image data) when the user initiates a ``scan'', which can be linear, from a script, or via the GEECS Python API. All devices are controlled by an instance of the same executable, which calls device specific code in a plugin architecture. GUIs are custom executables, but share the same communication code to receive data and send commands.}
\label{fig:GEECS}
\end{centering}
\end{figure*}

Many labs, in particular university-based systems, adapt or develop their own systems. The GEECS (generalized equipment and experiment control system) system provides an example of a home-grown control system developed over a number of years at the BELLA center which is open and modular.  Here, the GEECS team describe their system and considerations for others choosing to implement a home-grown control system. At the forefront, GEECS is designed to be manageable for researchers without a background in software engineering.\\

\paragraph{What is the BELLA Center?}
The Berkeley Lab Laser Accelerator (BELLA) Center focuses on the development and application of laser-plasma accelerators (LPAs). It houses four Ti:sapphire laser facilities. The highest peak power laser system is the BELLA PW \cite{Nakamura2017}, providing up to \unit[40]{J} on target in less than \unit[40]{fs} with a repetition rate of 1Hz. It has been utilized primarily for research on the high energy physics application of LPAs \cite{Gonsalves2019}, and more recently for ion acceleration studies and applications \cite{Hakimi2022} There are two \unit[60]{TW} systems operating at up to \unit[5]{Hz}, one which couples a sophisticated electron beamline including an undulator to its LPA \cite{Isono2021}, and another with two beamlines designed for MeV photon production via Thomson scattering \cite{Tsai2018}. And finally, the BELLA kHz LPA facility consists of a few-mJ few cycle laser system operating at kHz repetition rate, which can produce few-MeV electrons that are of interest in a variety of applications including medical and security. Each of these facilities is controlled and monitored by the generalized equipment and experiment control system (GEECS) developed in the BELLA Center, with typically dozens of computers, hundreds of devices, and thousands of process variables.

\paragraph{What is GEECS and why is it well suited for the BELLA Center?} GEECS is a complete software solution for control, monitoring, alarming and data logging of devices for process control and experimentation. For laser-plasma accelerators (LPAs) this means controlling laser, vacuum, timing, target, and diagnostic subsystems, and synchronizing data collection for each laser pulse. The GEECS architecture is shown in Figure \ref{fig:GEECS} and generally follows ANSI/IEEE-1471-2000 \cite{IEEE2000} as described in ``Documenting Software Architectures: Views and Beyond'' \cite{Clements2003}. The GEECS framework is scalable, distributed and modular, exhibiting a philosophy similar to well known platforms such as \hyperref[sec:epics]{EPICS}, with particular attention to making it easy to install, learn, use, customize, and expand. It was developed for the BELLA PW system \cite{Nakamura2017}, but has been the control system for all BELLA center facilities for over a decade.

The data acquisition system begins with the hardware devices such as cameras and motion controllers. GEECS aims to make the job of both the developer and user as easy as possible. For the developer, the software engineering effort for adding a new device type is minimized by limiting the effort to the code specific to that device, eliminating the need to write new user interfaces and configuration mechanisms for each new device type.

GEECS follows the object-oriented programming paradigm, which means that device types or ``classes'' inherit from the general device class as well as from each other, minimizing coding effort for adding a new device class. For example, this means the communication systems are automatically implemented for a new device class. Device classes also automatically have common methods such as ``acquire'', ``power off'' or ``load configuration''. The general device class is compiled into a single executable ``device.exe''. Any device launched uses this executable file, which then calls any device specific code required from source distribution folders.

New features can typically be added to individual devices without modifying the shared base executable. For the user, adding a new device is performed through a user interface that communicates with an SQL database describing the experimental configuration (the configuration database). Once a device is added to the database, it can be launched and controlled via a graphical user interface (GUI) named Master Control (MC). A few examples of the features of MC are that it can: start and shutdown all devices, computers, and GUIs associated with an experiment with a single click enabling rapid `switch-on' (or off) of an experiment; control any device on the system; perform scans of single or multiple variables; save configuration snapshots; and report system alarms. GEECS also includes GUIs for each type of device that makes configuration and viewing of data convenient. For example, the camera device type GUI allows for image viewing customization (such as colormaps, spatial smoothing, and perspective correction) and quick access to typical controls. In addition, users may develop custom GUI interfaces in a few minutes without the need for programming knowledge using LabVIEW and the GEECS publish/subscribe architecture, which is designed to reduce the dependencies between user interfaces and devices as much as possible. It should be noted that although the majority of code in GEECS is written in  LabVIEW, Python is increasingly being used to control and monitor GEECS devices, and has proven particularly useful for leveraging various machine learning tools.\\

\paragraph{Advantages of using LabVIEW.} Compared with traditional programming environments, the LabVIEW programming environment offers a number of advantages such as: reduced programming effort for many tasks such as interfacing with hardware and taking measurements, the same environment to develop desktop and real time applications, natural parallel processing, easy field-programmable gate array (FPGA) programming, and ease of data visualization. There are disadvantages also, including the need to purchase some of the plugins that reduce coding effort (e.g. NI Vision). The reason that we chose LabVIEW for the BELLA control system in 2009 is that the task of completing a full control system in a lower-level language is significant, and  abstraction in LabVIEW allows for rapid generation of code, whether it be the communication layer or hardware interface. This abstraction also allows for the control system to quickly adapt to new technologies.\\

\paragraph{Control System Architecture.}
Although we found certain advantages of using LabVIEW, the GEECS control system architecture is more important than the programming language used. By following standard practices such as ANSI/IEEE-1471-2000, coding effort is minimized and the system becomes far more adaptable. For example, the communication layer currently uses a mix of custom TCP and UDP, but could easily be re-written with abstractions to take advantage of a standardized message-passing library (such as ZeroMQ), since this layer is separate from others.
Our choice of object-oriented programming also brings advantages. Adding a new type of camera becomes straightforward since only the code specific to that type of camera needs to be written while other features are inherited.\\

\paragraph{Challenges and development path.}
The biggest challenge encountered has been the limited personnel effort available to develop the control system. Although a critical element to the success of modern LPA experiments, control systems have typically not been a priority. The choice to develop our own control system has allowed us to quickly add new device types with minimal effort, but at the same time it comes with some disadvantages. Although the framework does ensure some level of organization of the code, individual parts (especially the device specific code) is often rushed to get to a working state and never quite finished or polished. The most critical issue is that the documentation is insufficient. If the main developer ceases to work on this, it would be a challenge for a new developer to take over without help. User documentation is also a challenge, but this is one that can be addressed by encouraging users to contribute. And finally, the lifetime of GEECS is dependent on continued development and support of LabVIEW by National Instruments.\\

\paragraph{GEECS for others.}
Since the framework is based on standard best practices, the philosophy of the control system is similar to that of more widely used platforms. Those platforms could adapt some of the features of GEECS, especially the idea that a fully functional control system can be installed easily and configured quickly with simple GUIs, and that customization and new device classes can be made with little programming knowledge.

\subsubsection{Further Resources for LabVIEW}
The control system from the case study, GEECS, is open source and freely downloadable through the GEECS GitHub page \cite{GEECSgithub}. A link to a GEECS installation guide\cite{GEECSinstallation} is also found in this paper's references.  The NI Learning Center \cite{NILearn} provides professionally written resources for getting off the ground with LabVIEW. Also, NI has written a whitepaper on getting started with its Distributed Control and Automation Framework (DCAF) \cite{NIDCAF} (which is not used by GEECS).

\subsection{EPICS} \label{sec:epics}
\subsubsection{Overview of EPICS}

EPICS (the Experimental Physics and Industrial Control System) is an open-source framework for developing SCADA (Supervisory Control And Data Acquisition) systems. Originally developed at Argonne National Laboratory, it is based on applications communicating over the network using named PVs (process variables). As well as the core EPICS framework, the community has developed interfaces and drivers to support a wide variety of devices.

EPICS has depth of use and proven rigor in the particle physics, magnetic fusion, and astronomy community \cite{epicsprojects}.  EPICS is in use globally by many dozens of major scientific research facilities. Example facilities where EPICS is used today include high-energy-physics beamlines (e.g. Diamond Light Source, SLAC National Accelerator, Advanced Photon Source), astronomy and astrophysics observatories (e.g. W.M. Keck Observatory, LIGO), and magnetic confinement fusion facilities (e.g. KSTAR, ITER). The EPICS collaboration hosts annual meetings and code-athons welcome to anyone using EPICS \cite{epicsFAQ}.

One challenge of building EPICS systems for high-power laser facilities is a relatively small number of examples in our own community, especially for university-scale high-power laser facilities. Another challenge for members of our community in adopting EPICS is the need to develop device drivers for those scientific instruments not already in use in the broader EPICS community. Developing new drivers for devices in existing high-power laser systems may be a source of frustration for science teams, especially for small teams who are accustomed to relying on vendor-provided drivers or vendor-provided graphical user interfaces to control their scientific instruments. Fortunately, EPICS contains high-level tools to build new drivers and many abstractions that make this work easier. (For an example of a versatile device driver for cameras, see areaDetector \cite{areaDetector}).

There are often many ways to accomplish the same task in EPICS. For example, there are many alternate tools to display data, many alternate approaches to building data processing pipelines, two official communication protocols, several alternative techniques to remotely manage devices, and a variety of pathways to creating device drivers. A distributed community of EPICS developers over the past thirty years has created many interacting and overlapping solutions and toolsets. The number of choices among tools, and the deep and decentralized knowledge in the EPICS community, might be considered a long-term feature that also leads to a steep learning curve for new users.

EPICS is designed for control systems and control feedback loops involving process variables (e.g. valves, pumps, and pressure sensors), not specifically for scientific data acquisition (e.g. high-repetition-rate streak camera data). However, the latest version of EPICS, EPICS v7, includes a new network protocol (pVAccess) with data structures and network performance that are suitable for scientific data acquisition \cite{White2019}.

One major institution in the high-power laser community currently using EPICS is the Central Laser Facility at Rutherford Appleton Laboratory (RAL). We present their experience with EPICS in the following case study.

\subsubsection{Case Study: Central Laser Facility at RAL}

\begin{figure*}[htb]
\begin{centering}
\includegraphics[width = 17 cm]{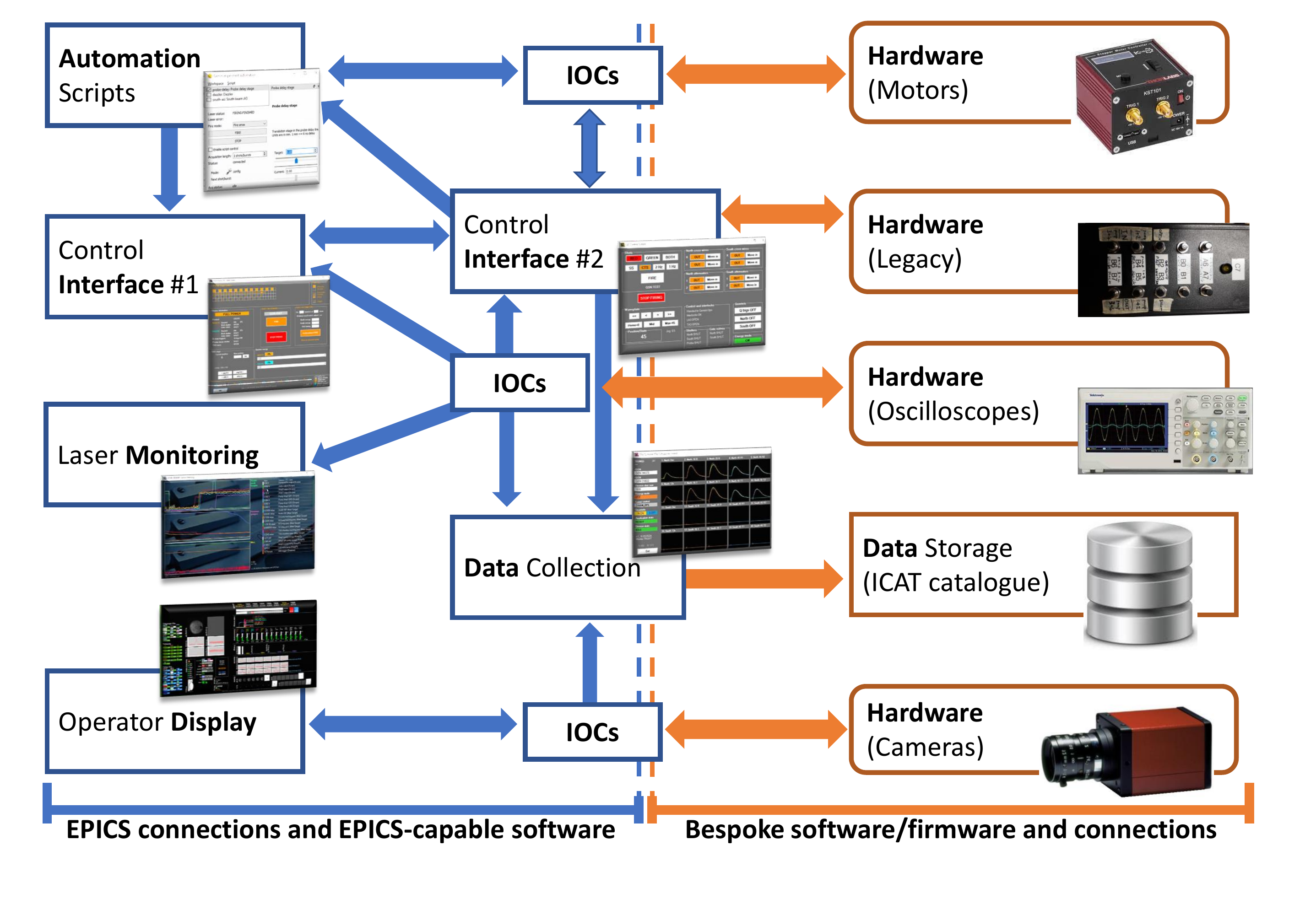}
\caption{Architecture of the Gemini Control System. EPICS Input/Output Controllers (IOCs) are software + hardware layers providing an abstraction between low-level device hardware and high-level control system software. Legacy hardware is connected into the system through a legacy interface, Control Interface \#2. A differentiation is made between EPICS-interfacing components and connections (in blue) and bespoke components and connections (in orange). Clients for device data include human control interfaces, informational laboratory displays, automation software, and archival data collectors.}
\label{fig:gemini}
\end{centering}
\end{figure*}

EPICS has been adopted fairly recently by the Central Laser Facility (CLF). It has been incorporated into the control system for the Gemini laser, was used to build the new control systems for HiLase and D100-X, and is being used to build the control system for the new Extreme Photonics Application Centre (EPAC facility) \cite{epac}.\\

\paragraph{EPICS in Gemini.} The Gemini Control System (Fig. \ref{fig:gemini}) was originally commissioned in 1997 for what was then the Astra laser facility.  Since then, it has undergone a number of upgrades, the most notable being the addition of twin Quantel lasers in 2009 when the facility was renamed Gemini. Since then, the Control System has been rewritten in .NET, and now uses EPICS.

There are several applications within the Control System suite:
\begin{enumerate}
\item The main Control System handles the orchestration of triggers, control of various devices, and (in conjunction with the interlock safety system) the hand-over of control from one laser area to another. 
\item Four Target Area Control Systems provide an interface to allow users of the facility to visualise the status of the beamline and to control devices appropriate to them. 
\item A Laser Area Control System allows the Gemini operators to visualise the status of the beamline from an operations point-of-view, and to set some of the main operational parameters.  
\end{enumerate}

These applications communicate using EPICS PVs. Each PV is a piece of data, usually small, that can be read and sometimes written by other applications. The main Control System makes available over 100 PVs. Examples include the laser operating (energy) mode, the status of wall shutters, and a 20 second countdown to the next shot. Applications monitor these PVs for various reasons, including changing device settings depending on laser energy mode and acquiring data on shot.

Other applications are used to control individual devices in the facility. These also communicate using EPICS PVs, and most of them host their own PVs. For example, the controller for a motorised stage may have a PV that can be written-to to move the stage, and another that can be read-from to find its current position. Controllers for thermometers, pressure gauges, oscilloscopes and other devices all make their data available through EPICS PVs in a similar way.

Any EPICS-aware software on the network can read and write EPICS PVs, so new instruments and applications can be added to the system without disrupting laser operations. Because writing to certain PVs could damage laser systems, Gemini operates within a private network and an EPICS Channel Access Gateway allows read-only access from other parts of site so operations can be monitored from the office without fear of inadvertent interference.

Diagnostic and metrology data from the laser are collected and stored in a data cataloguing system called ICAT provided by STFC’s Scientific Computing Department. The data are then made available to staff and users via a web interface called eCAT which also provides facilities to filter and analyse the data more fully, display traces and camera images in detail, and to download the data if required.\\

\paragraph{Experimental automation and data acquisition.} In contrast to the main operational diagnostics, those in the target area often change with each experiment so a more flexible data acquisition solution is needed. This has not been adapted to use EPICS. The main method used is a custom application called `Mirage'. Diagnostic software saves data to separate files – one file per laser pulse per diagnostic. Mirage then collects these files and saves them centrally, organised by diagnostic name, by run name, and by shot number. This makes it a simple operation to collect all the data for a single laser pulse. Mirage is also able to collect data from EPICS PVs. This is currently limited to environmental data, collected immediately before the laser pulse, but is invaluable for recording experimental settings.

Mirage offers several important features. It displays a summary of which diagnostics are acquiring data and allows users to perform a ``trigger test'', to ensure the system is working properly before a laser shot. It can also integrate with other systems. For example, by writing a short Python script, data can be analysed and the results plotted in real time.

A more recent feature is the ``experiment automation system''. This allows users or staff members to write Python scripts to control aspects of the experiment. These scripts can move (some) motors, adjust (some) laser parameters, and fire laser pulses. Although this system is still considered experimental, it has already been instrumental in some experiments requiring a high degree of automation \cite{Shalloo2020NC}.\\

\paragraph{EPICS in EPAC.} EPAC will build upon the experience and the lessons learned in Gemini while taking advantage of a dedicated team of software engineers to build more robust solutions that are more suitable for scaling to greater levels of complexity and higher data rates. It will also be able to take advantage of the wide variety of drivers developed by the EPICS community, as well as CLF’s experience developing EPICS-based control systems from the ground up for HiLase and D100X.

The control system for EPAC has some key differences from Gemini – notably device drivers are mostly implemented with standard EPICS frameworks (such as ``areaDetector'' for cameras and ``epics-motor'' for motion control). There are also some similarities, with most of the higher-level control system logic being implemented in .NET, which is more suitable for rapid prototyping of complex functionality than the native EPICS alternatives. A comprehensive user interface will be provided for laser operators, with a more limited version for facility users. These will be based on the Blazor framework, opening up the possibility of providing access through a web browser or on tablets and phones. This will be combined with user interfaces built with the widely used Control System Studio.

\paragraph{Data acquisition and management in EPAC.} While data acquisition will be performed by EPICS device drivers, a data management system is needed to collect and organise data into HDF5 files, which could be based on NeXus or any other standard format. This system is still under development, with key challenges being the high overall data rate (greater than \unit[1]{GB/s} expected) and the need to identify which laser pulse each piece of data is associated with.

The key technology for the data management system will be Apache Kafka, a distributed system for handling streams of data. Data will be sent to Kafka either by an EPICS-Kafka forwarder, or by plugins embedded into areaDetector drivers. Any system needing access to the data will be able to access a real-time stream, or recall it during a limited retention period. The latter may be particularly useful after a fault has occurred.

The aim is for a single file to contain all the data for a single sample or scan, as well as all the necessary metadata. Users will be able to access the data through DAaaS (Data Analysis as a Service, see \cref{scn:daaas}), and new systems will be provided for data archiving.

Notably, the greater use of EPICS and standardised interfaces within the EPAC control system should make it simpler for experiment automation systems to control various devices and system parameters. To take advantage of this, we intend to adopt Bluesky, an open-source framework for experiment automation.\\

\paragraph{Our experience with EPICS.} The use of EPICS was a considerable technological leap for the Gemini facility which, until 2018, was based mostly on UDP messaging.  Other beamlines within CLF had already started the move; so too had STFC’s ISIS, reassured that EPICS was championed by the Diamond Light Source on campus and other large facilities around the world.  
Like any software system EPICS has its quirks, and we were warned on many occasions that the learning curve was steep.  This has proved to be painfully true, but the advantages that it has brought in terms of flexibility and the ease of integration of new devices and applications have been considerable.

Another advantage is the wide community support. Many of the device drivers needed for EPAC already existed, substantially reducing the amount of custom code that must be written and supported. In addition, EPICS support is available for a wide variety of languages, including Python, Matlab, and LabView, allowing users to create custom applications to interact with the control system.

\subsubsection{Further Resources for EPICS}
The EPICS webpage provides a variety of official and community support resources \cite{epicssupport} for new members. Notably for members of our community who have never tried this control platform, EPICS can be explored on a virtual machine \cite{epicsVM} to sidestep a lengthy configuration, or integrated into a mockup physical control system built with low-cost Raspberry Pis \cite{sidekickepics}.

\subsection{Tango Controls}

\subsubsection{Overview of Tango Controls}
Tango Controls is a free open-source software toolkit \cite{tangodocs} for building object-oriented SCADA systems. It was originally developed at the European Synchrotron Radiation Facility (ESRF) \cite{TangoControlHistory} 20 years ago and has now been adopted by many scientific facilities, and in particular, telescopes, accelerators, light sources, and associated beamlines around the world. Tango Controls can either be used as the main toolkit for their control system, for a subsystem or together with commercially acquired systems in a local distributed network. Tango Controls relies on an active community \cite{TangoControlForum} of developers and users and is independent of operating system, supporting a core composed by libraries and API definitions in C++, Java and Python \cite{tango_docs_tango_core}.

Tango Controls aims to provide object-oriented programming for distributed heterogeneous systems. The Tango Controls software communication layer is based on Common Object Request Broker Architecture (CORBA)\cite{Corba}. CORBA brings a standard interface for all the objects and services available using an Interface Definition Language (IDL) \cite{InterfaceDefinitionLanguage}. An inter-operable object reference (IOR) identifies each object. One of the advantages is that it is not necessary to recompile when adding a new object. Since version 9 of Tango Controls, event-based communications use the zeroMQ library \cite{tango_docs_zeroMQ}.

\begin{figure}
    \centering
    \includegraphics[width=7cm]{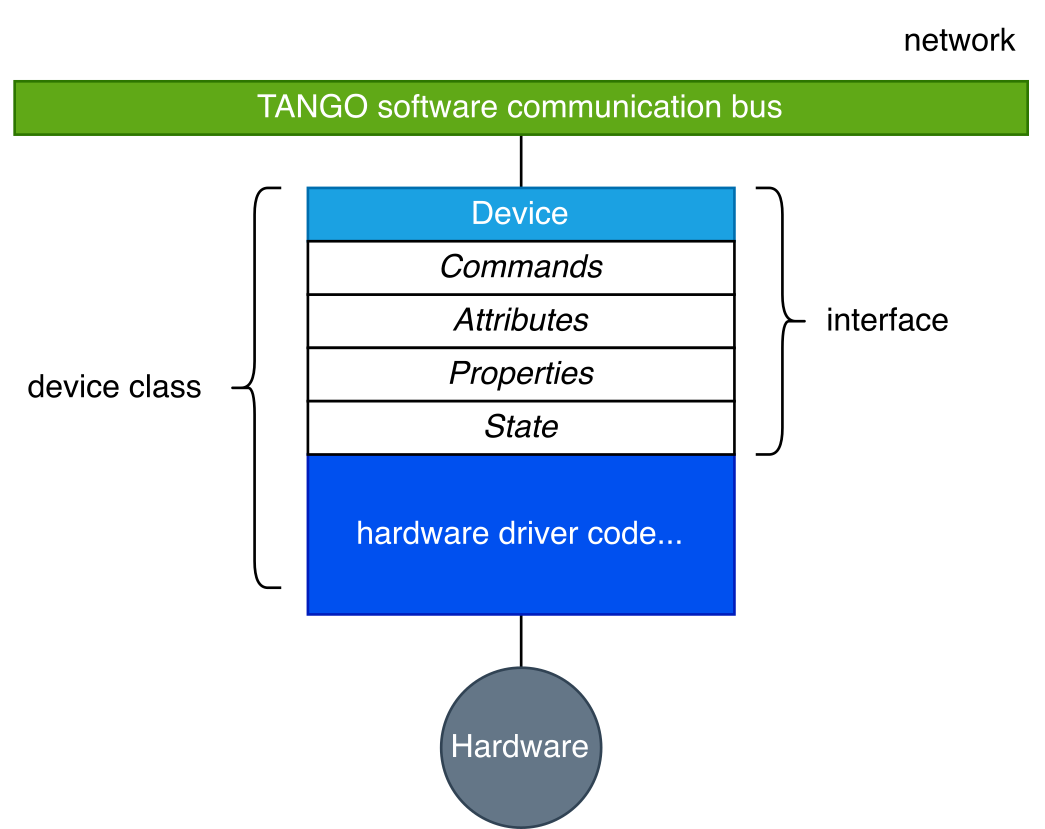}
    \caption{Illustration of the standard Tango device structure between the Tango Controls software communication bus and the hardware. The interface is common to all the device for all the device class and can be generated using the POGO tool \cite{pogo}. The hardware driver part code has to be written by the developer on the top of an existing driver, software development kit or communication protocol}
    \label{fig:tangoDeviceInterface}
\end{figure}

Before going further in the description of a representative system for the laser-plasma community, let us introduce some language elements specific to Tango. A \textit{device} is something that needs to be controlled. It can be equipment (e.g. a camera, a motor controller...),  a set of software functions or an ensemble of equipment (e.g. a deformable mirror and wavefront sensor). Then we have three tightly intertwined concepts.  The \textit{Tango class} defines the interface and the implementation of the device control as shown in figure \ref{fig:tangoDeviceInterface}: commands, attributes and properties. The commands act on the device (e.g. \textsc{ON, OFF, RESET}), the attributes set/get physical values of the device and properties are the configuration parameters (e.g. \textsc{IP address, port number...}). The \textit{Tango device} is an instance of a \textit{Tango class} giving access to the services of the class. The \textit{Tango device server} (DS) is the process in which one or more Tango classes are running, each one implementing a device. A hierarchical naming scheme is used for devices. Each device is identified by the Fully Qualified Domain Name \cite{fqdn} like \textsc{/domain/family/member}. 

The DS configuration is stored into the Tango database identified as the \textsc{tango\_host} server. The  device number and names for a Tango class are defined within the database while the Tango classes which are part of the DS are defined in the Tango database and in the source code of DS. 
The Tango database is associated to a special Tango device performing a centralized storage for the control system configuration parameters and for persistent data.
The Tango database is based on a MariaDB database engine. The Tango database is also the service for establishing connections (IOR) between client and server on the control system. So the minimum configuration Tango Controls system can be a computer unit running a MariaDB database server, the Tango database device and a DS.  On the same computer we can then run a DS for a camera (LIMA \cite{lima1}), a DS controlling a tip-tilt mirror and a DS being a software function steering a laser beam far field to a given position on the camera. 

\begin{figure}
    \centering
    \includegraphics[width=7cm]{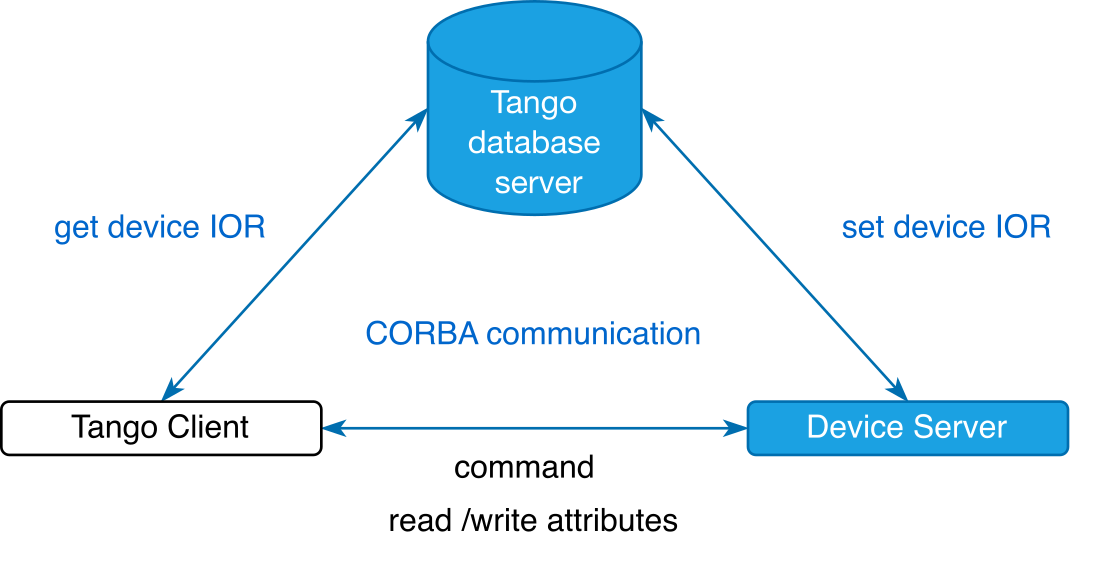}
    \caption{Illustration of the relation between client and device server and the Tango database running on the \textsc{tango\_host}. This simple system constitutes the minimal Tango configuration.}
    \label{fig:tangoCommunication}
\end{figure}

The device attributes are of various type\cite{tangodocs} and support three format: scalar, spectrum (1D array), and image (2D array). Attributes can be set in three access mode: read, write and read\&write. Each device attribute is defined by its properties which are fixed to five types in which one can find the data information, range value, and two essential properties for a distributed control system : \textit{Alarm} and \textit{Event} parameters.
When implementing Tango devices in a control system a hierarchy has to be set starting from the hardware with a low level device, then a device encompassing the relations between several low level devices and then higher level devices with processes and calculations on the sub devices attributes data.
The communication between device servers are of two main types. The client/server communication that can be asynchronous or synchronous or event-based with \textit{push/subscribe} method. The most used and simple asynchronous is the polling mode mechanism. It allows the Tango DS to decouple the real device from the clients requests. The Tango DS has got a specific polling threads, that can be configured for polling attributes or commands. The polling results are stored in a buffer with a configurable depth. This implementation helps to monitor the health status of the DS. Since Tango Controls 8 \textit{push/subscribe} event communication is available for the attributes. The main categories of pushed event are the change event (absolute or relative change on the attribute that can be configured), a periodic event, or archive event being a mix of the periodic event as the change of value is checked at the polling period. For further details, on alarms and logging specific Tango device, we invite the readers to have a look to the Tango documentation \cite{tangodocs}

Several systems in our high-power laser community, particularly European facilities\cite{tangoinstitutions}, are contributing to the Tango Controls platform. For example APOLLON \cite{TangoAPOLLON} ELI-ALPS\cite{TangoELIALPS}, ELI-BEAMLINES\cite{TangoELIBL} and CALA \cite{TangoCALA}. Below, we present a case study from one of them: the PALLAS project of the CNRS national institute for nuclear and particle physics (IN2P3) and host in Laboratoire de Physique des 2 infinis Irène Joliot-Curie (IJClab).

\subsubsection{Case Study: PALLAS}

\paragraph{What is the PALLAS project and why is Tango Controls well-suited for this work?} The PALLAS (prototypying accelerator based on laser-plasma technology) project hosted at Irène Joliot-Curie Lab (ICJLab) - also known as IJCLab - is developing a laser-plasma injector test facility with the goal of producing electron beams with \unit[200]{MeV}, \unit[30]{pC}, $<5\%$ energy spread and \unit[1]{mm-mrad} emittance at \unit[10]{Hz} with comparable stability and reliability to more conventional radio-frequency (RF) accelerators. The project has 3 main axes of investigation for the laser-plasma based injector: (1) advanced laser control, (2) development of plasma targetry and (3) development of a compact electron-beam-characterization-beamline for studies exploring accelerator staging and beam transport. A state-of-the-art control command and acquisition system is mandatory to achieve the optimizations and systematic studies on the reliability and stability of laser-plasma injectors. Based on the experiences of ICJlab groups on the ThomX project \cite{Thomx} and at neighboring facilities (ESRF, SOLEIL, APOLLON) who are actively involved in the development of Tango Controls, the choice of Tango Controls has been made for the PALLAS project.  A limitation for projects based on laser-plasma acceleration is the diversity of instruments and their use compared to conventional RF particle accelerators, especially on the whole laser driver control part. \\

\paragraph{How is Tango Controls integrated into PALLAS?} The global architecture of the PALLAS control command and acquisition system is as follows. The laser driver is a customized commercial 40 TW laser system running its own distributed control command system based on the commcercial ElliOOs libraries on a separated local network. A Tango gateway between the local distributed network of the laser system and the main PALLAS local network allows the control of pre-selected features of the laser system. The laser features currently accessible are, laser status, main shutter for laser firing, energy, spectrum and beam position and profile at the various stage of the laser system up to the last amplifier. All the hardware of the laser-plasma injector, including the laser transport and compression, are integrated under Tango Controls. The vacuum pumps, gauges, valves, and radiation safety system are controlled by a programmable logic controller (PLC) with a Modbus/TCP interface for Tango Controls supervisory control integration. The hardware for the laser-driven plasma accelerator are motors, area detectors, photo-diode detectors, magnet power supplies, digitizer and various data-acquisition (DAQ) cards. The total number of device servers for hardware control is currently 23 without including the device server for archiving operations and accelerator control or optimization. 

\begin{figure*}
    \begin{centering}
        \includegraphics[width = 17 cm]{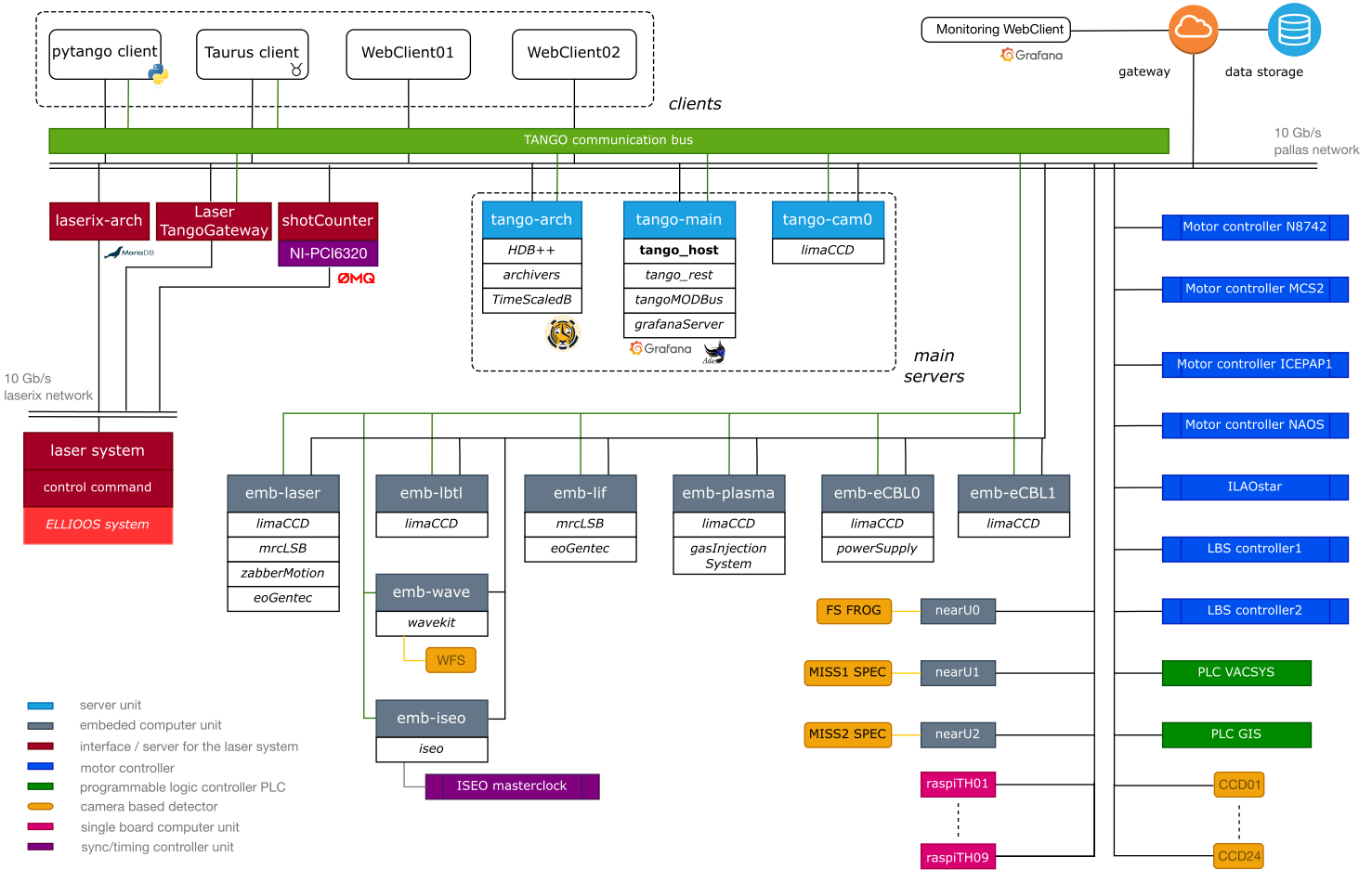}
        \caption{Schematic of the PALLAS control command and acquisition system. System elements are represented by colored rectangles (color legend in lower-left corner), with links between elements represented as colored lines. Green lines show Tango communication pathways, black lines indicate wired Ethernet connections, yellow lines show wired USB-3 connections, and gray lines indicate wired RS232 connections. For better readability, some hardware is not shown. The system can be divided into four parts: (1) the main Tango server units (light blue), the heart of the system running the Tango database, archiving system, and webservers; (2) the embedded computer units (light gray) running some Tango device servers (sometimes located adjacent to scientific hardware due to limited cable lengths for USB-3 or RS232 device connections); (3) the hardware motor controllers (dark blue), the PLCs (green), and the area detectors (dark yellow); (4) examples of clients: \textit{pyTango} and \textit{Taurus}, which need Tango Controls software installed on a client machine, and \textit{WebClients}, which require only a web browser and a connection to the local network.  Below the colored rectangle for each server unit or embedded computer unit, device servers running on that unit are indicated in italic font within white-filled rectangles. These are named \textit{LimaCCD, eoGENTENC, mrcLSB,...} and represent device servers openly developed and available for download \cite{TangoControlGit,PALLAS_gitlab}. The laser system has three connection points into the PALLAS local network: the \textit{laserix-arch} where laser data are stored, the \textit{shotCounter} which uniquely identifies laser shots, and the \textit{TangoGateway} running a device server which gives Tango access to certain laser controls. The laser system has its own separated network. As shown in upper right corner, some of the webservers can be accessed from either side of the PALLAS local network through a gateway, and data storage is done outside the local network for easier and broader data sharing.}
    \label{fig:pallasTango}
\end{centering}
\end{figure*}

The laser-driver local network is \unit[1]{Gb/s} network with special link at \unit[10]{Gb/s} between the main network switch and the data server \textit{laserix-arch}, while the accelerator network has a speed of \unit[10]{Gb/s} (see fig \ref{fig:pallasTango}). The laser system has its own data logging system based on a MariadB back-end. The event subscriber device-servers, or archivers, will be able to gather selected attributes and data from Tango DS with a unique shot identification and time-stamping given by the ``\textit{ShotCounter}'' system at \unit[10]{Hz}.

\begin{figure}
    \centering
    \includegraphics[width=8.5cm]{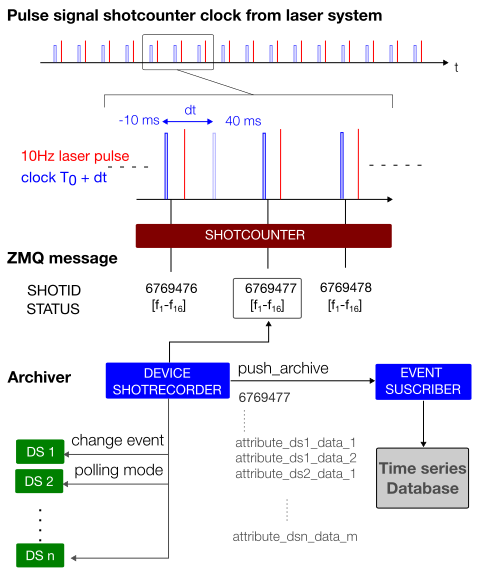}
    \caption{Timing and event generation system for the time and shot stamping of the data. A TTL signal can be tuned from -10 ms to 40 ms around each laser pulse at 10Hz. The shotCounter generated a unique 64-bit shot identification. The ZMQ message contains the SHOTID and STATUS of the 16 analog input of the NI-PCIe 6320 card. Archiving data is set by the \textit{SHOTRECORDER} device server (DS) and then an archiver (\textit{EVENT SUBSCRIBER} DS) stores the data to the target database.}
    \label{fig:pallasArchiving}
\end{figure}

On the accelerator part, the archiving system is currently under development and will be based on the archiving Tango system HDB++\cite{Bourtembourg2019} with a TimescaledB back-end and the  \textit{ShotCounter} system for event building based on a unique laser shot identification number (64-bit integer) and a configurable laser and system status. The \textit{ShotCounter}, supplied by the Amplitude Laser, is composed by an embedded PC and an NI PCIe 6320 card \cite{PCIcard} with 16 analog and digital I/O associated with a \unit[100]{MHz} digitization allowed by an NI-STF3 \cite{NISTC3} counter/timer. The embedded PC of the \textit{ShotCounter} runs a ZeroMQ \cite{zeroMQ} server configured in publisher mode which can distribute the event time-stamping and shot identification at up to \unit[100]{Hz} depending on the triggering configuration of hardware to both laser-driver and accelerator local network.  The 16 analog inputs of the \textit{ShotCounter} allow the recording of laser system states (energy level, amplifier injection, opening of the shutter to accelerator etc...) by applying threshold. All the logic is set within the \textit{SHOTRECORDER} DS as shown in figure Fig. \ref{fig:pallasArchiving}. The recording of attributes can be set depending on the DS communication type and then all the attribute data are pushed by the \textit{EVENT SUBSCRIBER} DS to the various Time series database.

The  data flux is expected to vary between \unit[50-250]{MB/s} depending on the \textit{EVENT SUBSCRIBER} DS configuration and the number of attributes to be stored. The database architecture is based on a high availability scheme with write to master database and read to replica for the accelerator beamline and for the laser driver system.

Tango Controls has a rich graphical client framework. Out of the box tools for Tango Controls include Astor\cite{astor} (a graphical tool allowing distant management of device servers without needing use of ssh), Jive\cite{jive} (a graphical tool for browsing \textsc{tango\_host} database and to configure devices), ATKPanel\cite{atkpanel} (allowing a default or developed GUI to control a device), Pogo (to generate devices), Taurus\cite{taurus} (to generate and develop a GUI), and Taranta\cite{tangoTaranta} or Waltz\cite{Waltz} (for web GUIs). For the PALLAS project, we foresee leveraging these various graphical clients to suit different needs (like web interfaces for broad accessibility on mobile, and Python for broad integration with other software). First, for the operational user interface of the laser system, laser transport, compression, focal spot characterization and optimization, and vacuum system control we plan to use a cross platform web-based dashboard with Ada webserver and plotly libraries \cite{Ada}\cite{plotly}, and API-REST Tango interface \cite{APIRest}. Second, for scan and optimization of the laser-plasma injector, Python-based scripts will be developed. Third, for monitoring of the whole system (infrastructure temperature, relative humidity, cooling water flow, gas pressure distribution, main laser-driver parameters, vacuum levels, radiation levels etc), we will use Grafana \cite{Grafana} on top of the archiving system. This will give us a distributed dashboard accessible from anywhere, including staff offices and various client posts within the laboratory.

\paragraph{Challenges.} One of the challenges for the laser-driven plasma accelerator is the extensive use of large area detectors and the necessity of data reduction/compression before the data is sent to the global network of the Tango Controls system. The laser-driver system uses 20 CCD cameras at 1 and \unit[10]{Hz}, distributed across two network switches, and two embedded PCs for image processing to determine the laser spot centroid, size and orientation. The accelerator part of the system, including laser transport, optical compression and focusing, as well as full characterization of the laser pulse in the interaction region, includes a total of 25 area detectors for laser beam diagnostics, electron beam profile and spectrometers.
Most of the laser diagnostics are being integrated into the Tango Controls system with standardization of the client-server configuration with REST-API protocol between embedded computer units, for example for laser diagnostics such as FROG\cite{Trebino1997} and 2D spectrometer. This configuration allows for 2D data-array processing, and retrieval software in the embedded computer units sending only reduced scalar data and processed image to the Tango device server on request.

\subsubsection{Further Resources for Tango Controls}
Tango Controls provides extensive official documentation\cite{tangodocs}, an active community forum\cite{TangoControlForum}, a searchable archive of a mailing list that preceded the community forum\cite{TangoMailingList}, and a Slack channel\cite{TangoSlack}. Each of these is accessible to new users. Notable for those wishing to quickly try out Tango Controls, a fully-configured virtual machine is available \cite{tangoinstall} for local installation or in cloud server services. Furthermore, the PALLAS project is participating as much as possible in Tango Controls development, favoring open-source annex API. All the device servers developed to support PALLAS are made available to the community within \textit{IN2P3}'s public GitLab repositories, under ``NOliProg/instrument'' \cite{PALLAS_gitlab}. 

\subsection{Intersections and Further Alternatives}
Facility-level approaches to laboratory control can utilize a mix of strategies and technologies while retaining their high-level organization. Planning for a mix of strategies and technologies may particularly benefit the high-power laser science community for two reasons. First, scientists in the high-power laser community often travel and repeat similar experiments at different facilities. When traveling, scientists often bring their own equipment to various facilities. Planning strategies in advance that integrate (rather than isolate) foreign subsystems will strengthen the scientific flexibility of a facility that accommodates traveling scientists. Second, planning for heterogeneity in technologies (rather than rigid uniformity) allows for quick adoption of advances made by other facilities without completely repeating their development effort.  For example, EPICS and Tango Controls devices could be graphically controlled by clients running LabVIEW or Taurus\cite{taurus}, or an EPICS client could be patched into the Tango Controls system. Network formats can be translated and adapted between control platforms. Custom Python control code, or low-level C++ instrument drivers, could be interfaced into any of a variety of control platforms. Further alternative control platforms exist, and we have given only a short overview of three that were discussed at a recent workshop on control systems for high-power laser systems\cite{controlworkshop2022}.

\section{Laboratory Data Management}

By coupling the increased access to multi-Hz laser systems with control systems that can maximise data acquisition rate, next-generation experiments will create orders of magnitude more output data than prior generations. Similarly, the explosion in available computing power has enabled greater simulations capability, permitting simulations with higher complexity to be performed within reasonable timescales.
In both cases, this data is accompanied by a multitude of metadata which is equally important to the accurate processing and long-term curating of the data. Without the development, or integration, of new approaches to data management and access, the advantages of this increased data rate cannot be realised. 

The traditional approach to high-power laser experimental data is very much rooted in ``single shot'' experiments.  During such experiments, groups of researchers might realistically expect to take a few hundred shots, with some small number of these shots repeats at the same point in the experimental parameters space to provide a measure of stability.  The data from all the experimental diagnostics would be stored locally totaling a few tens of Gbits, and copied to portable drives to be transported to different institutes for independent analysis.   Within the established approach to simulations, several low-dimension simulations may be run to optimise the simulation setup and scan the parameter space with a limited number of high-cost ``hero'' simulations performed at discrete points of interest.  Given the low volume of data it is possible for analysis to be strongly customized for individual shots / simulations.       
In the new systems, with terabytes to petabytes of data, this methodology of independent distributed copies of the data, and individually tailored analysis, may not be as effective at leveraging the facilities' capabilities in producing the highest impact science. The approach taken to store and interact with data may benefit from adaptation.

The very individual nature of experiments (and simulations) with varying diagnostics, file formats, file structures and analysis requirement also presents a challenge when attempting to combine or compare data between facilities or between different simulation codes.  
In order to increase efficiency, facilitate data comparison, and comply with open access requirements, it is necessary to determine ways in which data can be standardized and usefully accessed by the community.  These goals can be advanced by considering the re-use of diagnostic and analysis tools, standardized data formats and cloud-based processing. 

In this section, we describe approaches for addressing the challenges of data storage and access.

\subsection{Opportunities}
\subsubsection{Facilitating Data Access and Tooling Re-Use: F.A.I.R.}

The term F.A.I.R. is often used together with data and metadata. The acronym stands for Findable, Accessible, Interoperable and Reusable \cite{Wilkinson2016,gofair}.
In brief, the four letters describe how to enable automated access to and processing of data and metadata:
\begin{itemize}
\item{\textbf{F}indable: use a globally unique and persistent ID for each set, describe data with rich metadata, metadata must include that ID, and register the IDs at a searchable source.}
\item{\textbf{A}ccessible: Access to data and metadata via the ID by using an open and free protocol, also with authentication methods if necessary}
\item{\textbf{I}nteroperable: use a formal/accessible/shared language for knowledge representation, use vocabularies, cross-referencing}
\item{\textbf{R}eusable: provide also attributes for usage license and data provenance, establish and use community standards.}
\end{itemize}

F.A.I.R data and “Open Science” should not be confused despite the usual meaning of the terms “fair” or “fairness” referring to humanity and ethical behaviour. F.A.I.R data rather relates to machine-assisted research methods, and therefore also allows for restricted access in cases where this is demanded by policy. The F.A.I.R. principles basically address the needs for automated data processing but do not prescribe to which extent data and metadata must be publicly and freely available. These aspects are left to be determined by the owner. However, as soon as data accessibility is defined to follow the F.A.I.R. principles – which basically imply machine readability and compliance with a community-wide vocabulary – the step towards Open Science is relatively small.

A number of the technical requirements listed in the outline above have known solutions. For example, several methods exist to generate globally unique and persistent IDs, for example, DOI, URN or hdl (handle). These systems also register the respective IDs and provide access protocols. Similarly, several languages for knowledge representation exist, including Web Ontology Language 
\cite{webontlang}, ClearTalk, and CQL. Yet, the F.A.I.R. principles also suggest that domain-specific, community-wide accepted standards should be jointly defined. This comprises, in contrast to the aforementioned protocols, a consensus across the community about terms and their meaning, i.e. a vocabulary of the respective research field. Standardization is therefore an essential step towards F.A.I.R. and such standardization of data, vocabulary and behaviour has been adopted within other research communities for example the astrophysics community \cite{Molinaro2021,Allen2021}.

One significant example of adoption of the F.A.I.R. framework in the high-power laser community is at the Extreme Light Infrastructure (ELI). An ELI data policy document \cite{Weeks2022} speaks to rationale and implementation of F.A.I.R. principles in the context of high-power lasers, and can serve as motivation for further adoption in our community. A variety of general F.A.I.R. case studies can also be read online and are linked in our bibliography \cite{fairsfair, forbesfair, sheffieldfair}. 

\subsubsection{Definitive Data Sources and Shared Tools}

One opportunity in having a consistent laboratory approach to data management is to facilitate remote access to original data. Having an accessible, definitive source for original data can be helpful in avoiding mislabeled data or fragmented datasets that might result from passing data from person to person. Furthermore, facilities can provide access to contextual information (such as room temperature and humidity) which typically would not be requested by experimenters, except when hindsight leads to the need.

A data management approach might also involve non-local analysis of the remote datasets using shared analysis tools. Cloud software and cloud environments, configurable and shareable between stakeholders, could leverage distributed data access and high performance analysis on large datasets that a local computer with a local copy of data would not. Shared cloud resources for data analysis might lower the barrier to entry for new team members, and strengthen collaboration within and between teams.

Finally, definitive data sources and shared tooling have positive implications for scientific reproducibility. Others can re-trace steps in data analysis more easily, leading to more reproducible scientific analysis. Furthermore, with copious metadata and facility context, researchers can re-trace facility conditions and instrument configurations to reproduce scientific experiments with better fidelity.

\subsection{Challenges}
\subsubsection{Data Gatekeepers and Flexibility}
Formalizing approaches to data, and increasing accessibility, leads to questions of ``who sets our standards?'' and ``who has access?''

\paragraph{``Who has access?''} When each element of data is treated individually, this question is managed on a case-by-case basis by scientists and facility members. However, with automated structures in place for data management by the facility, data access questions are raised more explicitly and involve the facility to a higher degree. Choices of who does (and does not) have access to data may depend on the scientist preference or facility needs. It may be desired to share certain experimental data to all facility users. Or, it may be desired to limit data to just the facility team, or a single scientific team. One might also decide to limit data access to one group prior to a certain date. Understanding typical science data access patterns among facility stakeholders is important, so that facilities can set up access control tools which are technically capable of enabling these patterns. Best practice is to plan out access control \emph{prior to taking data} through conversations between the experiment team and the facility team. There is no one solution to access control for all experimental runs.

\paragraph{``Who sets and updates the standards?''} In the past, researchers working in isolation have been able to maintain their notes without oversight leading to huge variations in style and detail. This \emph{modus operandi} is not compatible with transparency and open access, and is especially not compatible with machine processing of metadata. While standards are necessary to support high data loads and open access, there is a new challenge of supporting heterogeneous instruments and data needs within a common framework -- and of bringing stakeholders to consensus on this framework.

Existing structure can make certain elements of creating new device data easier, but other elements less flexible. If data standards are too rigid, they can hinder scientific flexibility and may be ignored, weakening the effect of the standards. This particular challenge can be partly mitigated through use of open formats and flexible standards that allow for optional metadata fields and quick new definitions of formats.

\subsection{Case Studies}

We now discuss three concrete examples in our community to tackling the challenges of scientific data management: OpenPMD, DAaaS (Data Analysis as a Service), and the Michigan Cloud Computing Platform.

\subsubsection{OpenPMD, Software Tools and Metadata Standards based on F.A.I.R}
As an example for the laser-plasma domain, openPMD\cite{openpmdwebsite} is an open meta-standard initially designed for particle-mesh data originating from PIC simulations. Meta-standard means it defines how data shall be stored and organized, where “storing” includes a description of the data. Hence, in openPMD a number of properties are standardized like self-description and data structuring while others remain flexible. This has proven to be quite powerful since openPMD has recently been adopted \cite{openpmdcodes} by a number of widely used particle-in-cell (PIC) codes, including PIConGPU, Warp, WarpX, FBPIC, Wake-T, SimEx platform, LUME, ParaTAXIS, OSIRIS, UPIC-Emma, Sirepo/Warp PBA, Sirepo/Warp VND, HiPACE++, ACE3P, and CarpetX. Using a common format then allows for common analysis tools which gather in an ever-growing software ecosystem around openPMD.  However, while openPMD stores simulation output it does not store the input. Input files are of course much more dependent on the simulation package, hence the portability is not required from a technical point of view. However, for scientific reasons, inclusion of simulation input – ideally on an abstraction layer – would not only enable comparisons across simulation packages. It would furthermore increase the  comprehensibility of the output because it would be set into an intelligible context. This is a goal of PICMI \cite{PICMI}.

The situation for experimental data is even more complicated. Many diagnostics fielded in laser-plasma experiments record some kind of 2D data in the form of an image. This can be a laser beam profile or laser focus image, but also an image of a particle beam profile converted via a scintillator. More complex in processing are images of data from spectrometers, either dispersing photons or particles, where the spatial coordinates depend, among other parameters, on the energy of the particle/photon. As of now, openPMD is able to ingest the raw images with help of its CCDimage plugin \cite{openpmdccd}. Once stored, further openPMD tools can be used. However, detector properties and experiment geometry are crucial for analysis of raw data and must be included for compliance with F.A.I.R. This is the point where further standardization of domain-specific terms comes into play in order to achieve interoperability. The definition of such vocabularies must be done carefully with inclusion of the the community to ensure wide validity and acceptance.

In addition, experimental parameters such as target parameters, are decisive for interpretation of processed data. A problem emerging hereof is the flow of data and metadata: Sometimes, targets are characterized beforehand and often do not exist after the interaction. Broadening this, it can be expensive to automate annotation of everything of importance that occurs throughout the day in a laboratory. Examples include targets that are characterized entirely outside of the control system, or the unplanned manual adjustment of a mirror. Experiment geometry and detector parameters can be recorded all along an experimental campaign but may change in between shots. Raw data, in contrast, occurs only upon the shots. These are challenges for metadata, and requires an experimentalist to help keep metadata up-to-date. Including these asynchronous changes and data acquisitions into openPMD would be significant effort. However, for experimental data from the Photon and Neutron (PaN) science community there already exists a similar meta-standard, called the NeXus Data Format\cite{Konnecke2015}. It is possible that there might be sufficient analogies in data and metadata structure between laser-plasma experiments and PaN experiments to conceive a F.A.I.R. standard building upon openPMD and NeXus.

\subsubsection{Data Analysis as a Service at the Central Laser Facility} \label{scn:daaas}
Data Analysis as a Service (DAaaS) was originally developed for users of the ISIS neutron and muon source [https://www.isis.stfc.ac.uk/] to analyse the data produced by their experiments. By bringing together experimental data, compute resources, and specialised analysis software, it aims to remove the need for users to set up their own analysis environments. By providing everything needed for analysis as well as shared storage, DAaaS reduces the effort needed to add a new collaborator.

DAaaS is made up of virtual machines (VMs) running in the STFC cloud. Software is automatically preinstalled, and the VMs have access to shared storage containing experimental data. Access to this shared storage is controlled, so that a user can only see data for experiments in which they're involved. Users log into the system through a web browser and select which kind of ``workspace'' they require. This determines the compute resources (CPU, GPU, RAM) available to the VM, as well as the pre-installed analysis software. The system then obtains a freshly-installed VM from a pool. The user can then access the desktop of the VM (the XFCE desktop environment, running on Linux) through their web browser.

The VM provides access to a per-user persistent home directory, as well as read-only storage for experimental data and a read-write shared drive for analysis scripts and outputs. This allows users to collaborate with other users on the same experiment, by sharing their analysis work.

DAaaS is managed by STFC’s Scientific Computing Department, and has now been rolled out across the Central Laser Facility. Experimental data can be uploaded to the system through dedicated file servers. In Gemini, this can be done automatically by Mirage. This allows DAaaS to be used to analyse data immediately after it's been acquired.

Since being introduced, DAaaS has become popular among the CLF user community and has been used to build machine-learning models of experimental data \cite{Streeter2023HPLSE}, highlighting its suitability for data management and analysis.

\subsubsection{Michigan Cloud Computing Platform for Data Management}
The Michigan Cloud Computing Platform is a unified cloud platform for data management from numerical or physical experiments being developed by researchers at the University of Michigan. The platform seeks to enable FAIR principles of data management, as discussed previously, for small academic research teams, and therefore, to greatly lower the barrier-to-entry for performing Machine Learning (ML) on collected data. 

The platform is built on MLFlow \cite{mlflow}, an open source software developed for managing Machine Learning experiments, and the infrastructure-as-a-service tools provided by the public Cloud vendors e.g. Amazon Web Services.  There are two distinct usage patterns to the platform which cover the initial data gathering and subsequent analysis as follows:
\begin{itemize}
\item{\textbf{Usage Pattern 1} - At data collection time, the researcher is able to use a secure API to communicate (physical or numerical) experimental data to the server, which then organizes the data within the remote database and object storage. }
\item{\textbf{Usage Pattern 2} - Data can be accessed, analyzed, and downloaded securely via the web using the built-in GUI within MLFlow. The platform also has the ability to launch Jupyter Lab environments with flexible computing power (4-192 CPUs, 4GB-1TB of RAM) within the secured virtual private network. This enables researchers to perform data transformations and visualization from anywhere where they can access the web.}
\end{itemize}

We find that for our use case of storing , with redundancy, simulation and experimental data for a small academic group that amounts to approximately 25 TB, the storage costs are approximately \$500/month, and the server and database cost approximately \$200/month. If cloud data storage costs are prohibitive, MLFlow can also hook into other artifact stores (see \url{https://mlflow.org/docs/latest/tracking.html#artifact-stores}) like a typical NFS mount, or even a Hadoop Distributed File System. 

\paragraph{Advantages of MLFlow} The difference between local data management and cloud-based data management is primarily in the amount of work needed to build, deploy, and maintain the infrastructure, as well as in the ability to distribute access across varying geographic regions. The primary infrastructure of any data-management platform are servers, databases, and object storage mechanisms as well as the networking and security layers. 

The public Cloud providers have infrastructure-as-a-service capabilities for networking, security, servers, databases, and object storage. This takes the ``undifferentiated heavy-lifting'' of building, deploying, securing, and maintaining these constructs away from the scientific group. 

Now, the scientific group is primarily charged with customizing the data-structure and adapting their existing workflow. Using a pay-as-you-go approach also provides the scientific group with the freedom to upgrade/downgrade their hardware requirements as they wish. This helps mitigate possible restrictions from previous purchases and contracts that lock research groups into specific toolkits and hardware constraints. Additionally, it enables the scientific research group to be able to leverage the latest technology, e.g. GPU accelerators for training neural networks. 

Access to these services is provided through well-developed, performant, and stable software that is managed by the cloud companies. In fact, there exists a whole ecosystem of companies vying for market-share in this space alone. These well-developed software tools ensure the scientific group's accessibility of the data stored on the cloud. In a final note on data accessibility, the public cloud vendors also invest heavily in redundancy in their systems. In fact, they advertise that their systems, when configured properly, only lose data very infrequently (better than one part in a billion) and rarely have outages in service. 

It is also important to consider the data security needs of an academic research group. Using a cloud-based approach enables the research group to leverage the significant investments made by the cloud companies in order to court high-data-security industries like the financial services industry. This ensures that the data-security needs of a small-medium scale scientific research group are met with relatively little overhead by simply just using the standard workflows and permissioning schemes prescribed by the cloud provider. Data, by default, is not publicly accessible, but rather, is made available to access by the person or team given the authority to manage the permissions. 

It is likely that larger labs and companies that use cloud-based-solutions for (computational or physical) scientific data-management have internal solutions that do something similar to our proposed solution but broader in scope. However, at this time, it is not clear to us whether there are easily accessible, alternative cloud solutions for scientific researchers that can be maintained by 1 or 2 cloud engineers. This is why we propose our solution. The primary advantages are that it is relatively low-budget, is easy-to-implement and maintain in an academic research setting, and can be scaled up or down, resource-wise, as is necessary.

A clear disadvantage is using MLFlow as a layer between the user and the database. This is not the most performant implementation but is much more user-friendly. As the data-management effort scales up and teams can be tasked with maintaining the platform, it is likely that a more performant and scalable implementation does not have the MLFlow layer as an intermediary and uses an alternative for accessibility such as a GUI.

\paragraph{Challenges of MLFlow} The primary challenge we have encountered so far is to convince researchers to adapt their existing workflow and use the platform in order to derive benefits from it over the medium to long term. 

Historically, the typical researcher was incentivized to extract what they aimed for out of the data and discard the data. While the platform discussed here decreases the friction of collecting and using data over a medium-to-long term, it still remains a challenge to incentivize researchers and decrease the friction with regards to adoption in the short-term. 

This challenge is partially addressed by the draw of big-data and ML. As the high-power Laser community develops novel methods and performs research using ML, the data management strategy for each individual research group becomes increasingly important. This is because ML requires data management principles in two flavors. 

First, and most obviously, running the ML algorithms, at scale, on a dataset requires the dataset to be organized, cleaned, and easily accessible. This is a primary principle behind the development of this platform. 

The second flavor of data management required for ML comes from a more subtle point. Each instantiation of the machine learning algorithm over a dataset can be considered a numerical experiment. In this case, much like with a simulation, it is also important to track the different parameters provided to the machine learning experiments and be able to easily correlate those to the outcome of the numerical experiment.

For these two reasons, the draw of developing ML techniques in high-power laser research provides an incentive to the small-group researcher to improve data-management principles in the short term.

At small-to-medium scale facilities, the folks running the facility may find it useful to collect data over many experiments, months, and years in order to better understand the performance and behavior over time. This can also drive an interest in having a platform that can collect and organize this data.

Another challenge is having the right expertise in place to be able to deploy such a platform for a small academic research group or facility. While cloud platforms offer a GUI that can help launch the right servers, networking tools, databases etc., it can be quite cumbersome to repeat this task as well as debug it in case something goes awry. To address this, we have developed the platform using an Infrastructure-as-Code approach. This effectively means that there is a code-base that, when run, deploys the entire platform platform readily into different cloud environments, e.g. each belonging to a different team or group. This reduces the time required to replicate the deployment of the platform from days to minutes.

\paragraph{Community input} One of the central beliefs behind developing our platform is that over the lifetime of a (numerical or physical) experiment, data generated by researchers can often get ``siloed''. With the growing applications of algorithms that can provide data-driven insights, the ``siloed'' data represents potentially useful information that is effectively lost. 

To be able to best utilize these data in downstream, machine-learning-related, tasks, it would behoove researchers to approach data collection and management as a longer-term effort than just the short time spent performing and analyzing the numerical or physical experiment. Our platform hopefully lowers the barrier-to-entry for such an effort.

We are currently using the platform as a test-bed for managing experimental data for a user facility, for managing simulation and machine learning data within an academic research group \cite{joglekar_2022}, and for machine learning experiment management.

\paragraph{How can I try MLFLow?} Most current open-source machine-learning experiment managers can run locally on a personal computer where the object and database storage are also hosted locally.  This configuration can be installed as a Python package in the typical manner (using the Python package management software ``pip'') and executed from the command line. However, the experiment manager system can also be deployed in a more scalable manner on to a remote server with long term object storage and a database. In this case, it is preferable to build a Docker container to deploy on the remote server. At the time of this writing, we prefer using MLFlow as our experiment manager, but there are many other similar open-source experiment managers that serve a similar purpose. 

\section{Conclusion}

The high-power laser community is at a critical moment in laying down new, long-lasting digital infrastructure for its facilities. Through this manuscript, which addresses a pressing community need, we hope to facilitate open and efficient future collaboration. We aim to spearhead this by supporting knowledge-sharing of the work being done to develop both control systems and data management to meet the needs of the next generation of high-power laser experiments. A distributed networked control system can facilitate laboratory-wide operational speeds and closed-loop approaches which humans cannot achieve. A consistent approach to managing data can increase data accessibility to scientists and external partners, increase reliability of metadata, and increase re-usability of data-analysis software. In this manuscript, we explored considerations for practical facility-level decision-making in these areas, and we highlighted several specific control systems and approaches to data management from our community. By taking steps now to communicate and synchronize, our community can access the benefits, and mitigate the challenges, of our next-generation-facility digital infrastructure.

\bibliography{main.bbl}

\begin{thebibliography}{100}

\bibitem{Corde2013RMP}
S.~Corde, K.~Ta~Phuoc, G.~Lambert, R.~Fitour, V.~Malka, A.~Rousse, A.~Beck, and
  E.~Lefebvre.
\newblock Femtosecond x rays from laser-plasma accelerators.
\newblock {\em Rev. Mod. Phys.}, 85:1--48, Jan 2013.

\bibitem{Couperus2017NC}
J.~P. Couperus, R.~Pausch, A.~Köhler, O.~Zarini, J.~M. Krämer, M.~Garten,
  A.~Huebl, R.~Gebhardt, U.~Helbig, S.~Bock, K.~Zeil, A.~Debus, M.~Bussmann,
  U.~Schramm, and A.~Irman.
\newblock Demonstration of a beam loaded nanocoulomb-class laser wakefield
  accelerator.
\newblock {\em Nature Communications}, 8(1):487, 2017.
\newblock Number: 1 Publisher: Nature Publishing Group.

\bibitem{Gotzfried2020PRX}
J.~Götzfried, A.~Döpp, M.~F. Gilljohann, F.~M. Foerster, H.~Ding,
  S.~Schindler, G.~Schilling, A.~Buck, L.~Veisz, and S.~Karsch.
\newblock Physics of high-charge electron beams in laser-plasma wakefields.
\newblock {\em Physical Review X}, 10(4):041015, 2020.
\newblock Publisher: American Physical Society.

\bibitem{Foerster2022PRX}
F.~M. Foerster, A.~Döpp, F.~Haberstroh, K.~v. Grafenstein, D.~Campbell, Y.-Y.
  Chang, S.~Corde, J.~P. Couperus~Cabadağ, A.~Debus, M.~F. Gilljohann, A.~F.
  Habib, T.~Heinemann, B.~Hidding, A.~Irman, F.~Irshad, A.~Knetsch,
  O.~Kononenko, A.~Martinez de~la Ossa, A.~Nutter, R.~Pausch, G.~Schilling,
  A.~Schletter, S.~Schöbel, U.~Schramm, E.~Travac, P.~Ufer, and S.~Karsch.
\newblock Stable and high-quality electron beams from staged laser and plasma
  wakefield accelerators.
\newblock {\em Physical Review X}, 12(4):041016, 2022.
\newblock Publisher: American Physical Society.

\bibitem{Leemans2014PRL}
W.~P. Leemans, A.~J. Gonsalves, H.-S. Mao, K.~Nakamura, C.~Benedetti, C.~B.
  Schroeder, Cs. Tóth, J.~Daniels, D.~E. Mittelberger, S.~S. Bulanov, J.-L.
  Vay, C.~G.~R. Geddes, and E.~Esarey.
\newblock Multi-{GeV} electron beams from capillary-discharge-guided
  subpetawatt laser pulses in the self-trapping regime.
\newblock {\em Physical Review Letters}, 113(24):245002, 2014.
\newblock Publisher: American Physical Society.

\bibitem{Gonsalves2019PRL}
A.~J. Gonsalves, K.~Nakamura, J.~Daniels, C.~Benedetti, C.~Pieronek, T.~C.~H.
  de~Raadt, S.~Steinke, J.~H. Bin, S.~S. Bulanov, J.~van Tilborg, C.~G.~R.
  Geddes, C.~B. Schroeder, Cs. T\'oth, E.~Esarey, K.~Swanson, L.~Fan-Chiang,
  G.~Bagdasarov, N.~Bobrova, V.~Gasilov, G.~Korn, P.~Sasorov, and W.~P.
  Leemans.
\newblock Petawatt laser guiding and electron beam acceleration to 8 gev in a
  laser-heated capillary discharge waveguide.
\newblock {\em Phys. Rev. Lett.}, 122:084801, Feb 2019.

\bibitem{Kneip2009PRL}
S.~Kneip, S.~R. Nagel, S.~F. Martins, S.~P.~D. Mangles, C.~Bellei, O.~Chekhlov,
  R.~J. Clarke, N.~Delerue, E.~J. Divall, G.~Doucas, K.~Ertel, F.~Fiuza,
  R.~Fonseca, P.~Foster, S.~J. Hawkes, C.~J. Hooker, K.~Krushelnick, W.~B.
  Mori, C.~A.~J. Palmer, K.~Ta Phuoc, P.~P. Rajeev, J.~Schreiber, M.~J.~V.
  Streeter, D.~Urner, J.~Vieira, L.~O. Silva, and Z.~Najmudin.
\newblock Near-gev acceleration of electrons by a nonlinear plasma wave driven
  by a self-guided laser pulse.
\newblock {\em Phys. Rev. Lett.}, 103:035002, Jul 2009.

\bibitem{Kim2021AS}
Hyung~Taek Kim, Vishwa~Bandhu Pathak, Calin~Ioan Hojbota, Mohammad Mirzaie,
  Ki~Hong Pae, Chul~Min Kim, Jin~Woo Yoon, Jae~Hee Sung, and Seong~Ku Lee.
\newblock Multi-{GeV} laser wakefield electron acceleration with {PW} lasers.
\newblock {\em Applied Sciences}, 11(13):5831, 2021.
\newblock Number: 13 Publisher: Multidisciplinary Digital Publishing Institute.

\bibitem{Badziak2018JP}
J.~Badziak.
\newblock Laser-driven ion acceleration: methods, challenges and prospects.
\newblock {\em Journal of Physics: Conference Series}, 959(1):012001, 2018.
\newblock Publisher: {IOP} Publishing.

\bibitem{Schreiber2016RSI}
J.~Schreiber, P.~R. Bolton, and K.~Parodi.
\newblock Invited review article: “hands-on” laser-driven ion acceleration:
  A primer for laser-driven source development and potential applications.
\newblock {\em Review of Scientific Instruments}, 87(7):071101, 2016.
\newblock Publisher: American Institute of Physics.

\bibitem{Alejo2015}
A.~Alejo, H.~Ahmed, A.~Green, S.~R. Mirfayzi, M.~Borghesi, and S.~Kar.
\newblock Recent advances in laser-driven neutron sources.
\newblock {\em Il nuovo cimento C}, 38(6):1--7, 2015.
\newblock Number: 6 Publisher: Societa italiana di fisica.

\bibitem{Albert2014PPCF}
F.~Albert, A.~G.~R. Thomas, S.~P.~D. Mangles, S.~Banerjee, S.~Corde, A.~Flacco,
  M.~Litos, D.~Neely, J.~Vieira, Z.~Najmudin, R.~Bingham, C.~Joshi, and
  T.~Katsouleas.
\newblock Laser wakefield accelerator based light sources: potential
  applications and requirements.
\newblock {\em Plasma Physics and Controlled Fusion}, 56(8):084015, 2014.
\newblock Publisher: {IOP} Publishing.

\bibitem{Wenz2015NC}
J.~Wenz, S.~Schleede, K.~Khrennikov, M.~Bech, P.~Thibault, M.~Heigoldt,
  F.~Pfeiffer, and S.~Karsch.
\newblock Quantitative x-ray phase-contrast microtomography from a compact
  laser-driven betatron source.
\newblock {\em Nature Communications}, 6(1):7568, 2015.
\newblock Number: 1 Publisher: Nature Publishing Group.

\bibitem{Cole2015SR}
J.~M. Cole, J.~C. Wood, N.~C. Lopes, K.~Poder, R.~L. Abel, S.~Alatabi, J.~S.J.
  Bryant, A.~Jin, S.~Kneip, K.~Mecseki, D.~R. Symes, S.~P.D. Mangles, and
  Z.~Najmudin.
\newblock Laser-wakefield accelerators as hard x-ray sources for {{3D}} medical
  imaging of human bone.
\newblock {\em Scientific Reports}, 5(1):13244, October 2015.

\bibitem{Dopp2018Optica}
A.~Döpp, L.~Hehn, J.~Götzfried, J.~Wenz, M.~Gilljohann, H.~Ding,
  S.~Schindler, F.~Pfeiffer, and S.~Karsch.
\newblock Quick x-ray microtomography using a laser-driven betatron source.
\newblock {\em Optica}, 5(2):199--203, 2018.
\newblock Publisher: Optica Publishing Group.

\bibitem{Wang2021Nature}
Wentao Wang, Ke~Feng, Lintong Ke, Changhai Yu, Yi~Xu, Rong Qi, Yu~Chen, Zhiyong
  Qin, Zhijun Zhang, Ming Fang, Jiaqi Liu, Kangnan Jiang, Hao Wang, Cheng Wang,
  Xiaojun Yang, Fenxiang Wu, Yuxin Leng, Jiansheng Liu, Ruxin Li, and Zhizhan
  Xu.
\newblock Free-electron lasing at 27 nanometres based on a laser wakefield
  accelerator.
\newblock {\em Nature}, 595(7868):516--520, 2021.
\newblock Number: 7868 Publisher: Nature Publishing Group.

\bibitem{Pompili2022Nature}
R.~Pompili, D.~Alesini, M.~P. Anania, S.~Arjmand, M.~Behtouei, M.~Bellaveglia,
  A.~Biagioni, B.~Buonomo, F.~Cardelli, M.~Carpanese, E.~Chiadroni, A.~Cianchi,
  G.~Costa, A.~Del~Dotto, M.~Del~Giorno, F.~Dipace, A.~Doria, F.~Filippi,
  M.~Galletti, L.~Giannessi, A.~Giribono, P.~Iovine, V.~Lollo, A.~Mostacci,
  F.~Nguyen, M.~Opromolla, E.~Di~Palma, L.~Pellegrino, A.~Petralia,
  V.~Petrillo, L.~Piersanti, G.~Di~Pirro, S.~Romeo, A.~R. Rossi, J.~Scifo,
  A.~Selce, V.~Shpakov, A.~Stella, C.~Vaccarezza, F.~Villa, A.~Zigler, and
  M.~Ferrario.
\newblock Free-electron lasing with compact beam-driven plasma wakefield
  accelerator.
\newblock {\em Nature}, 605(7911):659--662, 2022.
\newblock Number: 7911 Publisher: Nature Publishing Group.

\bibitem{Graydon2022NP}
Oliver Graydon.
\newblock The race for wakefield-driven {FELs}.
\newblock {\em Nature Photonics}, 16(11):750--751, 2022.
\newblock Number: 11 Publisher: Nature Publishing Group.

\bibitem{Esplen2020}
Nolan Esplen, Marc~S. Mendonca, and Magdalena Bazalova-Carter.
\newblock Physics and biology of ultrahigh dose-rate ({FLASH}) radiotherapy: a
  topical review.
\newblock {\em Physics in Medicine \& Biology}, 65(23):23TR03, 2020.
\newblock Publisher: {IOP} Publishing.

\bibitem{Chaudhary2021}
Pankaj Chaudhary, Giuliana Milluzzo, Hamad Ahmed, Boris Odlozilik, Aaron
  {McMurray}, Kevin~M. Prise, and Marco Borghesi.
\newblock Radiobiology experiments with ultra-high dose rate laser-driven
  protons: Methodology and state-of-the-art.
\newblock {\em Frontiers in Physics}, 9, 2021.

\bibitem{Zhang2018OE}
Jinlong Zhang, Hongfei Jiao, Bin Ma, Zhanshan Wang, and Xinbin Cheng.
\newblock Laser-induced damage of nodular defects in dielectric multilayer
  coatings.
\newblock {\em Optical Engineering}, 57(12):121909, 2018.
\newblock Publisher: {SPIE}.

\bibitem{Zhaoyang2022}
Zhaoyang Li, Yuxin Leng, and Ruxin Li.
\newblock Further development of the short-pulse petawatt laser: Trends,
  technologies, and bottlenecks.
\newblock {\em Laser \& Photonics Reviews}, page 2100705, 2022.

\bibitem{Prencipe2017}
I.~Prencipe, J.~Fuchs, S.~Pascarelli, D.~W. Schumacher, R.~B. Stephens, N.~B.
  Alexander, R.~Briggs, M.~Büscher, M.~O. Cernaianu, A.~Choukourov, M.~De
  Marco, A.~Erbe, J.~Fassbender, G.~Fiquet, P.~Fitzsimmons, C.~Gheorghiu,
  J.~Hund, L.~G. Huang, M.~Harmand, N.~J. Hartley, A.~Irman, T.~Kluge,
  Z.~Konopkova, S.~Kraft, D.~Kraus, V.~Leca, D.~Margarone, J.~Metzkes,
  K.~Nagai, W.~Nazarov, P.~Lutoslawski, D.~Papp, M.~Passoni, A.~Pelka, J.~P.
  Perin, J.~Schulz, M.~Smid, C.~Spindloe, S.~Steinke, R.~Torchio, C.~Vass,
  T.~Wiste, R.~Zaffino, K.~Zeil, T.~Tschentscher, U.~Schramm, and T.~E. Cowan.
\newblock Targets for high repetition rate laser facilities: needs, challenges
  and perspectives.
\newblock {\em High Power Laser Science and Engineering}, 5:e17, 2017.
\newblock Publisher: Cambridge University Press.

\bibitem{Chagovets2021}
Timofej Chagovets, Stanislav Stanček, Lorenzo Giuffrida, Andriy Velyhan,
  Maksym Tryus, Filip Grepl, Valeriia Istokskaia, Vasiliki Kantarelou, Tuomas
  Wiste, Juan~Carlos Hernandez~Martin, Francesco Schillaci, and Daniele
  Margarone.
\newblock Automation of target delivery and diagnostic systems for high
  repetition rate laser-plasma acceleration.
\newblock {\em Applied Sciences}, 11(4):1680, 2021.
\newblock Number: 4 Publisher: Multidisciplinary Digital Publishing Institute.

\bibitem{George2019}
K.~M. George, J.~T. Morrison, S.~Feister, G.~K. Ngirmang, J.~R. Smith, A.~J.
  Klim, J.~Snyder, D.~Austin, W.~Erbsen, K.~D. Frische, J.~Nees, C.~Orban,
  E.~A. Chowdhury, and W.~M. Roquemore.
\newblock High-repetition-rate ( {kHz}) targets and optics from liquid
  microjets for high-intensity laser–plasma interactions.
\newblock {\em High Power Laser Science and Engineering}, 7:e50, 2019.
\newblock Publisher: Cambridge University Press.

\bibitem{Treffert2022}
F.~Treffert, G.~D. Glenn, H.-G.~J. Chou, C.~Crissman, C.~B. Curry, D.~P.
  {DePonte}, F.~Fiuza, N.~J. Hartley, B.~Ofori-Okai, M.~Roth, S.~H. Glenzer,
  and M.~Gauthier.
\newblock Ambient-temperature liquid jet targets for high-repetition-rate {HED}
  discovery science.
\newblock {\em Physics of Plasmas}, 29(12):123105, 2022.
\newblock Publisher: American Institute of Physics.

\bibitem{Kraft2018}
Stephan~D. Kraft, Lieselotte Obst, Josefine Metzkes-Ng, Hans-Peter Schlenvoigt,
  Karl Zeil, Sylvain Michaux, Denis Chatain, Jean-Paul Perin, Sophia~N. Chen,
  Julien Fuchs, Maxence Gauthier, Thomas~E. Cowan, and Ulrich Schramm.
\newblock First demonstration of multi-{MeV} proton acceleration from a
  cryogenic hydrogen ribbon target.
\newblock {\em Plasma Physics and Controlled Fusion}, 60(4):044010, 2018.
\newblock Publisher: {IOP} Publishing.

\bibitem{Puyuelo2019}
P.~Puyuelo-Valdes, J.-L. Henares, F.~Hannachi, T.~Ceccotti, J.~Domange,
  M.~Ehret, E.~D'Humieres, L.~Lancia, J.-R. Marquès, J.~Santos, and
  M.~Tarisien.
\newblock Laser driven ion acceleration in high-density gas jets.
\newblock In {\em Laser Acceleration of Electrons, Protons, and Ions V}, volume
  11037, pages 14--21. {SPIE}, 2019.

\bibitem{Oertel2019}
John~A. Oertel, Cris~William Barnes, Michael Demkowicz, Gilliss Dyer, Mike
  Farrell, Martin Green, Ross Muenchausen, Abbass Nikroo, and Irene Prencipe.
\newblock Adaptive sample preparation and target fabrication for
  high-throughput materials science.

\bibitem{Grace2021PPCF}
E.~S. Grace, T.~Ma, Z.~Guang, R.~A. Simpson, G.~G. Scott, D.~Mariscal,
  B.~Stuart, and R.~Trebino.
\newblock Rapid retrieval of first-order spatiotemporal distortions for
  ultrashort laser pulses.
\newblock {\em Plasma Physics and Controlled Fusion}, 63(12):124005, 2021.
\newblock Publisher: {IOP} Publishing.

\bibitem{Obst2018PPCF}
L.~Obst, J.~Metzkes-Ng, S.~Bock, G.~E. Cochran, T.~E. Cowan, T.~Oksenhendler,
  P.~L. Poole, I.~Prencipe, M.~Rehwald, C.~Rödel, H.-P. Schlenvoigt,
  U.~Schramm, D.~W. Schumacher, T.~Ziegler, and K.~Zeil.
\newblock On-shot characterization of single plasma mirror temporal contrast
  improvement.
\newblock {\em Plasma Physics and Controlled Fusion}, 60(5):054007, 2018.
\newblock Publisher: {IOP} Publishing.

\bibitem{Waxer2018}
L.~J. Waxer, C.~Dorrer, A.~Kalb, E.~M. Hill, and W.~Bittle.
\newblock Single-shot temporal characterization of kilojoule-level, picosecond
  pulses on {OMEGA} {EP}.
\newblock In {\em Frontiers in Ultrafast Optics: Biomedical, Scientific, and
  Industrial Applications {XVIII}}, volume 10522, pages 98--106. {SPIE}, 2018.

\bibitem{Downer2018RMP}
M.~C. Downer, R.~Zgadzaj, A.~Debus, U.~Schramm, and M.~C. Kaluza.
\newblock Diagnostics for plasma-based electron accelerators.
\newblock {\em Reviews of Modern Physics}, 90(3):035002, 2018.
\newblock Publisher: American Physical Society.

\bibitem{Hatfield2021NAT}
Peter~W. Hatfield, Jim~A. Gaffney, Gemma~J. Anderson, Suzanne Ali, Luca
  Antonelli, Suzan Başeğmez~du Pree, Jonathan Citrin, Marta Fajardo, Patrick
  Knapp, Brendan Kettle, Bogdan Kustowski, Michael~J. {MacDonald}, Derek
  Mariscal, Madison~E. Martin, Taisuke Nagayama, Charlotte A.~J. Palmer, J.~Luc
  Peterson, Steven Rose, J.~J. Ruby, Carl Shneider, Matt J.~V. Streeter, Will
  Trickey, and Ben Williams.
\newblock The data-driven future of high-energy-density physics.
\newblock {\em Nature}, 593(7859):351--361, 2021.
\newblock Number: 7859 Publisher: Nature Publishing Group.

\bibitem{Dopp2022arxiv}
Andreas Döpp, Christoph Eberle, Sunny Howard, Faran Irshad, Jinpu Lin, and
  Matthew Streeter.
\newblock Data-driven science and machine learning methods in laser-plasma
  physics, 2022.

\bibitem{Loughran2023HPLSE}
B.~Loughran, M.~J.~V. Streeter, H.~Ahmed, S.~Astbury, M.~Balcazar, M.~Borghesi,
  N.~Bourgeois, C.~B. Curry, S.~J.~D. Dann, S.~DiIorio, N.~P. Dover,
  T.~Dzelzanis, O.~C. Ettlinger, M.~Gauthier, L.~Giuffrida, G.~D. Glenn, S.~H.
  Glenzer, J.~S. Green, R.~J. Gray, G.~S. Hicks, C.~Hyland, V.~Istokskaia,
  M.~King, D.~Margarone, O.~McCusker, P.~McKenna, Z.~Najmudin,
  C.~Parisua{\~n}a, P.~Parsons, C.~Spindloe, D.~R. Symes, A.~G.~R. Thomas,
  F.~Treffert, N.~Xu, and C.~a.~J. Palmer.
\newblock Automated control and optimisation of laser driven ion acceleration.
\newblock {\em High Power Laser Science and Engineering}, pages 1--11, March
  2023.

\bibitem{Jalas2021PRL}
Sören Jalas, Manuel Kirchen, Philipp Messner, Paul Winkler, Lars Hübner,
  Julian Dirkwinkel, Matthias Schnepp, Remi Lehe, and Andreas~R. Maier.
\newblock Bayesian optimization of a laser-plasma accelerator.
\newblock {\em Physical Review Letters}, 126(10):104801, 2021.
\newblock Publisher: American Physical Society.

\bibitem{Shalloo2020NAT}
R.~J. Shalloo, S.~J.~D. Dann, J.-N. Gruse, C.~I.~D. Underwood, A.~F. Antoine,
  C.~Arran, M.~Backhouse, C.~D. Baird, M.~D. Balcazar, N.~Bourgeois, J.~A.
  Cardarelli, P.~Hatfield, J.~Kang, K.~Krushelnick, S.~P.~D. Mangles, C.~D.
  Murphy, N.~Lu, J.~Osterhoff, K.~Põder, P.~P. Rajeev, C.~P. Ridgers,
  S.~Rozario, M.~P. Selwood, A.~J. Shahani, D.~R. Symes, A.~G.~R. Thomas,
  C.~Thornton, Z.~Najmudin, and M.~J.~V. Streeter.
\newblock Automation and control of laser wakefield accelerators using bayesian
  optimization.
\newblock {\em Nature Communications}, 11(1):6355, 2020.
\newblock Number: 1 Publisher: Nature Publishing Group.

\bibitem{Dann2019PRAB}
S.~J.~D. Dann, C.~D. Baird, N.~Bourgeois, O.~Chekhlov, S.~Eardley, C.~D.
  Gregory, J.-N. Gruse, J.~Hah, D.~Hazra, S.~J. Hawkes, C.~J. Hooker,
  K.~Krushelnick, S.~P.~D. Mangles, V.~A. Marshall, C.~D. Murphy, Z.~Najmudin,
  J.~A. Nees, J.~Osterhoff, B.~Parry, P.~Pourmoussavi, S.~V. Rahul, P.~P.
  Rajeev, S.~Rozario, J.~D.~E. Scott, R.~A. Smith, E.~Springate, Y.~Tang,
  S.~Tata, A.~G.~R. Thomas, C.~Thornton, D.~R. Symes, and M.~J.~V. Streeter.
\newblock Laser wakefield acceleration with active feedback at 5 hz.
\newblock {\em Physical Review Accelerators and Beams}, 22(4):041303, 2019.
\newblock Publisher: American Physical Society.

\bibitem{Spindloe2018preface}
Christopher Spindloe, Yuji Fukuda, Paul Fitzsimmons, Kai Du, and Colin Danson.
\newblock Review of {HPLSE} special issue on target fabrication.
\newblock {\em High Power Laser Science and Engineering}, 6:e13, 2018.
\newblock Publisher: Cambridge University Press.

\bibitem{Heuer2022preface}
P.~V. Heuer, S.~Feister, D.~B. Schaeffer, and H.~G. Rinderknecht.
\newblock Preface to special topic: The high repetition rate frontier in
  high-energy-density physics.
\newblock {\em Physics of Plasmas}, 29(11):110401, 2022.
\newblock Publisher: American Institute of Physics.

\bibitem{Ge2017IS}
Xiaohua Ge, Fuwen Yang, and Qing-Long Han.
\newblock Distributed networked control systems: A brief overview.
\newblock {\em Information Sciences}, 380:117--131, 2017.

\bibitem{Schramm2017}
U.~Schramm, M.~Bussmann, A.~Irman, M.~Siebold, K.~Zeil, D.~Albach, C.~Bernert,
  S.~Bock, F.~Brack, J.~Branco, J.~P. Couperus, T.~E. Cowan, A.~Debus,
  C.~Eisenmann, M.~Garten, R.~Gebhardt, S.~Grams, U.~Helbig, A.~Huebl,
  T.~Kluge, A.~Köhler, J.~M. Krämer, S.~Kraft, F.~Kroll, M.~Kuntzsch,
  U.~Lehnert, M.~Loeser, J.~Metzkes, P.~Michel, L.~Obst, R.~Pausch, M.~Rehwald,
  R.~Sauerbrey, H.~P. Schlenvoigt, K.~Steiniger, and O.~Zarini.
\newblock First results with the novel petawatt laser acceleration facility in
  dresden.
\newblock {\em Journal of Physics: Conference Series}, 874(1):012028, 2017.
\newblock Publisher: {IOP} Publishing.

\bibitem{Albach2019}
D.~Albach, M.~Loeser, M.~Siebold, and U.~Schramm.
\newblock Performance demonstration of the {PEnELOPE} main amplifier {HEPA} i
  using broadband nanosecond pulses.
\newblock {\em High Power Laser Science and Engineering}, 7:e1, 2019.
\newblock Publisher: Cambridge University Press.

\bibitem{Siebold2013}
Mathias Siebold, Fabian Roeser, Markus Loeser, Daniel Albach, and Ulrich
  Schramm.
\newblock {PEnELOPE}: a high peak-power diode-pumped laser system for
  laser-plasma experiments.
\newblock In {\em High-Power, High-Energy, and High-Intensity Laser Technology;
  and Research Using Extreme Light: Entering New Frontiers with Petawatt-Class
  Lasers}, volume 8780, pages 31--44. {SPIE}, 2013.

\bibitem{Michel2016}
Peter~Dr Michel.
\newblock Elbe center for high-power radiation sources.
\newblock {\em Journal of large-scale research facilities JLSRF}, 2:A39--A39,
  2016.

\bibitem{llinspector}
Invigon.
\newblock Industrial laser software (laser light inspector).
\newblock \url{https://invigon.de/en/products/camera-software}.
\newblock [Online; accessed 21st December 2022].

\bibitem{ffsync}
Zenju.
\newblock Freefilesync: Open source file synchronization software.
\newblock \url{https://freefilesync.org}.
\newblock [Online; accessed 21st December 2022].

\bibitem{MediaWiki}
MediaWiki.
\newblock Mediawiki website.
\newblock \url{https://www.mediawiki.org}.
\newblock [Online; accessed 12th November 2022].

\bibitem{scicat}
SciCat.
\newblock Scicat project.
\newblock \url{https://scicatproject.github.io}.
\newblock [Online; accessed 21st December 2022].

\bibitem{MongoDB}
MongoDB.
\newblock Mongodb.
\newblock \url{https://www.mongodb.com}.
\newblock [Online; accessed 12th November 2022].

\bibitem{Grafana}
Grafana.
\newblock Grafana.
\newblock \url{https://grafana.com/oss/grafana}.
\newblock [Online; accessed 12th November 2022].

\bibitem{Jenni2003ATLAS}
Peter Jenni, Marzio Nessi, Markus Nordberg, and Kenway Smith.
\newblock {ATLAS} high-level trigger, data-acquisition and controls: Technical
  design report.

\bibitem{Berger2008ATLAS}
N.~Berger, T.~Bold, T.~Eifert, G.~Fischer, S.~George, J.~Haller, A.~Hoecker,
  J.~Masik, M.~Z. Nedden, V.~P. Reale, C.~Risler, C.~Schiavi, J.~Stelzer, and
  X.~Wu.
\newblock The {ATLAS} high level trigger steering.
\newblock 119(2):022013.

\bibitem{Collab2020ATLAS}
The~{ATLAS} collaboration.
\newblock Operation of the {ATLAS} trigger system in run 2.
\newblock 15(10):P10004.

\bibitem{Thayer2017LCLSII}
Jana~B. Thayer, Gabriella Carini, Wilko Kroeger, Chris O'Grady, Amedeo Perazzo,
  Murali Shankar, and Matt Weaver.
\newblock Building a data system for {LCLS}-{II}.
\newblock In {\em 2017 {IEEE} Nuclear Science Symposium and Medical Imaging
  Conference ({NSS}/{MIC})}, pages 1--4.
\newblock {ISSN}: 2577-0829.

\bibitem{Gessner2022facetii}
S.~J. Gessner.
\newblock The {FACET}-{II} data acquisition system.

\bibitem{Rouse2007book}
William~B Rouse.
\newblock {\em People and organizations: Explorations of human-centered
  design}.
\newblock John Wiley \& Sons, 2007.

\bibitem{Hidvegi2011EuroXFEL}
Attila Hidvegi, Patrick Gessler, Kay Rehlich, and Christian Bohm.
\newblock Timing and triggering system prototype for the {XFEL} project.
\newblock 58(4):1852--1856.
\newblock Conference Name: {IEEE} Transactions on Nuclear Science.

\bibitem{Gaget2022SARAF}
Alexis Gaget.
\newblock {MRF} timing system design at {SARAF}.
\newblock pages 912--915. {JACOW} Publishing, Geneva, Switzerland.
\newblock {ISSN}: 2226-0358.

\bibitem{icalepcs2019}
Proceedings of the 17th international conference on accelerator and large
  experimental physics control systems: Timing and synchronization.
\newblock \url{https://accelconf.web.cern.ch/icalepcs2019/html/clas013.htm}.
\newblock [Online; accessed 3rd May 2023].

\bibitem{icalepcs2021}
Proceedings of the 18th international conference on accelerator and large
  experimental physics control systems: Timing systems, synchronization and
  real-time applications.
\newblock \url{https://accelconf.web.cern.ch/icalepcs2021/html/clas014.htm}.
\newblock [Online; accessed 3rd May 2023].

\bibitem{TI2012PTP}
Texas Instrument.
\newblock An-1728 ieee 1588 precision time protocol time synchronization
  performance.
\newblock {\em Application Report, April}, 2012, 2013.

\bibitem{Krzyzanowski2021website}
Paul Krzyzanowski.
\newblock Precision time protocol.
\newblock \url{https://people.cs.rutgers.edu/~pxk/417/notes/ptp.html}, 2021.
\newblock [Online; accessed 3rd May 2023].

\bibitem{zeroMQwiki}
ZeroMQ.
\newblock Zeromq wiki.
\newblock \url{http://wiki.zeromq.org}.
\newblock [Online; accessed 3rd May 2023].

\bibitem{MRFpricelist}
Micro-Research~Finland Oy.
\newblock Product price list.
\newblock \url{http://www.mrf.fi/index.php/product-price-list}.
\newblock [Online; accessed 3rd May 2023].

\bibitem{MicrosemiProducts}
Microsemi Corporation.
\newblock Products \& services: Timing \& synchronization.
\newblock
  \url{https://www.microsemi.com/product-directory/3425-timing-synchronization}.
\newblock [Online; accessed 3rd May 2023].

\bibitem{Nakamura2017}
Kei Nakamura, Hann-Shin Mao, Anthony~J. Gonsalves, Henri Vincenti, Daniel~E.
  Mittelberger, Joost Daniels, Arturo Magana, Csaba Toth, and Wim~P. Leemans.
\newblock Diagnostics, control and performance parameters for the {BELLA} high
  repetition rate petawatt class laser.
\newblock {\em {IEEE} Journal of Quantum Electronics}, 53(4):1--21, 2017.
\newblock Conference Name: {IEEE} Journal of Quantum Electronics.

\bibitem{Gonsalves2019}
A.~J. Gonsalves, K.~Nakamura, J.~Daniels, C.~Benedetti, C.~Pieronek, T.~C.~H.
  de~Raadt, S.~Steinke, J.~H. Bin, S.~S. Bulanov, J.~van Tilborg, C.~G.~R.
  Geddes, C.~B. Schroeder, Cs. Tóth, E.~Esarey, K.~Swanson, L.~Fan-Chiang,
  G.~Bagdasarov, N.~Bobrova, V.~Gasilov, G.~Korn, P.~Sasorov, and W.~P.
  Leemans.
\newblock Petawatt laser guiding and electron beam acceleration to 8 {GeV} in a
  laser-heated capillary discharge waveguide.
\newblock {\em Physical Review Letters}, 122(8):084801, 2019.
\newblock Publisher: American Physical Society.

\bibitem{Hakimi2022}
Sahel Hakimi, Lieselotte Obst-Huebl, Axel Huebl, Kei Nakamura, Stepan~S.
  Bulanov, Sven Steinke, Wim~P. Leemans, Zachary Kober, Tobias~M. Ostermayr,
  Thomas Schenkel, Anthony~J. Gonsalves, Jean-Luc Vay, Jeroen van Tilborg,
  Csaba Toth, Carl~B. Schroeder, Eric Esarey, and Cameron G.~R. Geddes.
\newblock Laser–solid interaction studies enabled by the new capabilities of
  the {iP}2 {BELLA} {PW} beamline.
\newblock {\em Physics of Plasmas}, 29(8):083102, 2022.
\newblock Publisher: American Institute of Physics.

\bibitem{Isono2021}
Fumika Isono, Jeroen~van Tilborg, Samuel~K. Barber, Joseph Natal, Curtis
  Berger, Hai-En Tsai, Tobias Ostermayr, Anthony Gonsalves, Cameron Geddes, and
  Eric Esarey.
\newblock High-power non-perturbative laser delivery diagnostics at the final
  focus of 100-{TW}-class laser pulses.
\newblock {\em High Power Laser Science and Engineering}, 9:e25, 2021.
\newblock Publisher: Cambridge University Press.

\bibitem{Tsai2018}
H.-E. Tsai, C.~G.~R. Geddes, T.~Ostermavr, G.~O. Muñoz, J.~van Tilborg, S.~K.
  Barber, F.~Isono, H.-S. Mao, K.~K. Swanson, R.~Lehe, A.~J. Gonsalves,
  K.~Nakamura, Cs. Toth, C.~B. Schroeder, E.~Esarey, and W.~P. Leemans.
\newblock Laser-plasma - accelerator-driven quasi - monoenergetic {MeV} thomson
  photon source and laser facility.
\newblock In {\em 2018 {IEEE} Advanced Accelerator Concepts Workshop ({AAC})},
  pages 1--5, 2018.

\bibitem{IEEE2000}
{IEEE} recommended practice for architectural description for
  software-intensive systems.
\newblock {\em {IEEE} Std 1471-2000}, pages 1--30, 2000.
\newblock Conference Name: {IEEE} Std 1471-2000.

\bibitem{Clements2003}
Paul Clements, David Garlan, Reed Little, Robert Nord, and Judith Stafford.
\newblock Documenting software architectures: views and beyond.
\newblock In {\em 25th International Conference on Software Engineering, 2003.
  Proceedings.}, pages 740--741. {IEEE}, 2003.

\bibitem{GEECSgithub}
GEECS Developers.
\newblock Geecs github page.
\newblock \url{https://github.com/GEECS-BELLA}.
\newblock [Online; accessed 13th January 2023].

\bibitem{GEECSinstallation}
GEECS Developers.
\newblock Geecs installation guide.
\newblock \url{https://sites.google.com/a/lbl.gov/geecs/installation}.
\newblock [Online; accessed 13th January 2023].

\bibitem{NILearn}
NI.
\newblock Ni learning center.
\newblock \url{https://learn.ni.com}.
\newblock [Online; accessed 13th January 2023].

\bibitem{NIDCAF}
NI.
\newblock Introduction to the distributed control and automation framework
  (dcaf).
\newblock
  \url{https://www.ni.com/en-us/innovations/white-papers/18/introduction-to-the-distributed-control-and-automation-framework.html}.
\newblock [Online; accessed 13th January 2023].

\bibitem{epicsprojects}
EPICS Collaboration.
\newblock Epics official webpage: Projects.
\newblock \url{https://epics-controls.org/epics-users/projects}.
\newblock [Online; accessed 20th December 2022].

\bibitem{epicsFAQ}
EPICS Collaboration.
\newblock Epics official webpage: Frequently asked questions.
\newblock
  \url{https://epics-controls.org/resources-and-support/documents/epics-faq}.
\newblock [Online; accessed 20th December 2022].

\bibitem{areaDetector}
Mark Rivers.
\newblock areadetector.
\newblock \url{https://areadetector.github.io/areaDetector}.
\newblock [Online; accessed 22nd April 2023].

\bibitem{White2019}
Greg White, Murali Shankar, Andrew~Nicholas Johnson, Mark~Lloyd Rivers, Guobao
  Shen, Sinisa Veseli, Kunal Shroff, Matej Sekoranja, Tom Cobb, Tim Korhonen,
  David~Gareth Hickin, Heinz Junkes, Martin~Gregor Konrad, Ralph Lange,
  Steven~M. Hartman, Kay-Uwe Kasemir, Matthew Pearson, Klemen Vodopivec,
  Leo~Bob Dalesio, Michael Davidsaver, Dirk Zimoch, and Martin~Richard Kraimer.
\newblock The {EPICS} software framework moves from controls to physics.
\newblock In {\em 10th International Particle Accelerator Conference
  ({IPAC}2019)}. {JACoW}, 2019.

\bibitem{epac}
CLF.
\newblock Clf introducing the extreme photonics application centre.
\newblock \url{https://www.clf.stfc.ac.uk/Pages/EPAC-introduction-page.aspx}.
\newblock [Online; accessed 22nd December 2022].

\bibitem{Shalloo2020NC}
R.~J. Shalloo, S.~J.~D. Dann, J.-N. Gruse, C.~I.~D. Underwood, A.~F. Antoine,
  C.~Arran, M.~Backhouse, C.~D. Baird, M.~D. Balcazar, N.~Bourgeois, J.~A.
  Cardarelli, P.~Hatfield, J.~Kang, K.~Krushelnick, S.~P.~D. Mangles, C.~D.
  Murphy, N.~Lu, J.~Osterhoff, K.~P{\~o}der, P.~P. Rajeev, C.~P. Ridgers,
  S.~Rozario, M.~P. Selwood, A.~J. Shahani, D.~R. Symes, A.~G.~R. Thomas,
  C.~Thornton, Z.~Najmudin, and M.~J.~V. Streeter.
\newblock Automation and control of laser wakefield accelerators using bayesian
  optimization.
\newblock {\em Nature Communications}, 11(1):6355, Dec 2020.

\bibitem{epicssupport}
EPICS Collaboration.
\newblock Epics official webpage: Support.
\newblock \url{https://epics-controls.org/support}.
\newblock [Online; accessed 13th January 2023].

\bibitem{epicsVM}
EPICS Collaboration.
\newblock Epics uspas 2019 virtual machine.
\newblock
  \url{https://controlssoftware.sns.ornl.gov/training/2019_USPAS/#prep}.
\newblock [Online; accessed 13th January 2023].

\bibitem{sidekickepics}
Scott Feister.
\newblock Sidekick epics.
\newblock \url{https://sfeister.github.io/sidekick-epics-docs}.
\newblock [Online; accessed 13th January 2023].

\bibitem{tangodocs}
Tango~Controls Collaboration.
\newblock Tango controls official documentation.
\newblock \url{https://tango-controls.readthedocs.io/en/latest}.
\newblock [Online; accessed 13th January 2023].

\bibitem{TangoControlHistory}
Tango~Controls Collaboration.
\newblock Tango controls history.
\newblock \url{https://www.tango-controls.org/about-us/#History}.
\newblock [Online; accessed 22nd February 2023].

\bibitem{TangoControlForum}
Tango~Controls Collaboration.
\newblock Tango controls collaboration forum.
\newblock \url{https://www.tango-controls.org/community/forum}.
\newblock [Online; accessed 22nd February 2023].

\bibitem{tango_docs_tango_core}
Tango~Controls Collaboration.
\newblock Glossary entry, tango core: Tango controls official documentation.
\newblock
  \url{https://tango-controls.readthedocs.io/en/latest/reference/glossary.html"}.
\newblock [Online; accessed 22nd April 2023].

\bibitem{Corba}
Alan~LaMont Pope.
\newblock {\em The CORBA Reference Guide: Understanding the Common Object
  Request Broker Architecture}.
\newblock Addison-Wesley Longman Publishing Co., Inc., USA, 1998.

\bibitem{InterfaceDefinitionLanguage}
OMG Standards~Development Organization.
\newblock Interface definition language (idl) specification, version 4.2.
\newblock \url{https://www.omg.org/spec/IDL/4.2}.
\newblock [Online; accessed 22nd April 2023].

\bibitem{tango_docs_zeroMQ}
Tango~Controls Collaboration.
\newblock Overview of tango controls: Tango controls official documentation.
\newblock
  \url{https://tango-controls.readthedocs.io/en/latest/overview/overview.html#what-is-tango-controls}.
\newblock [Online; accessed 22nd April 2023].

\bibitem{pogo}
Tango~Controls Collaboration.
\newblock Tango controls code generator (pogo).
\newblock
  \url{https://tango-controls.readthedocs.io/en/latest/tools-and-extensions/built-in/pogo/index.html}.
\newblock [Online; accessed 22nd April 2023].

\bibitem{fqdn}
P.~V. Mockapetris.
\newblock Rfc1035: Domain names - implementation and specification, 1987.

\bibitem{lima1}
Tango Controls~Collaboration ESRF.
\newblock Lima : Library for image acquisition.
\newblock \url{https://gitlab.esrf.fr/limagroup/lima}, 2022.

\bibitem{tangoinstitutions}
Tango~Controls Collaboration.
\newblock Tango controls official webpage: Institutions.
\newblock \url{https://www.tango-controls.org/partners/institutions}.
\newblock [Online; accessed 13th January 2023].

\bibitem{TangoAPOLLON}
Paillard J.-L. et~al.
\newblock {Apollon project status}.
\newblock In {\em {32nd Tango Collaboration Meeting}}, 2018.

\bibitem{TangoELIALPS}
Schrettner L. et~al.
\newblock {ELI-ALPS Control System Status }.
\newblock In {\em {32nd Tango Collaboration Meeting}}, 2018.

\bibitem{TangoELIBL}
Bastl P. et~al.
\newblock {ELI Beamlines Control system status}.
\newblock In {\em {32nd Tango Collaboration Meeting}}, 2018.

\bibitem{TangoCALA}
Nils Weiße, Leonard Doyle, Johannes Gebhard, Felix Balling, Florian Schweiger,
  Florian Haberstroh, Laura~D. Geulig, Jinpu Lin, Faran Irshad, Jannik
  Esslinger, and et~al.
\newblock Tango controls and data pipeline for petawatt laser experiments.
\newblock {\em High Power Laser Science and Engineering}, page 1–8, 2023.

\bibitem{Thomx}
The thomx ics source.
\newblock {\em Physics Open}, 5:100051, 2020.

\bibitem{TangoControlGit}
Tango~Controls Collaboration.
\newblock Tango controls gitlab / documentation.
\newblock \url{https://gitlab.com/tango-controls}.
\newblock [Online; accessed 22nd February 2023].

\bibitem{PALLAS_gitlab}
PALLAS.
\newblock Pallas gitlab - noliprog/instrument.
\newblock \url{https://gitlab.in2p3.fr/noliprog/instrument}.
\newblock [Online; accessed 12th November 2022].

\bibitem{Bourtembourg2019}
Reynald Bourtembourg et~al.
\newblock {Pushing the Limits of Tango Archiving System using PostgreSQL and
  Time Series Databases}.
\newblock In {\em {17th International Conference on Accelerator and Large
  Experimental Physics Control Systems}}, page WEPHA020, 2020.

\bibitem{PCIcard}
NI.
\newblock Pcie-6320 multifunction i/o device.
\newblock \url{https://www.ni.com/en-gb/support/model.pcie-6320.html}.
\newblock [Online; accessed 12th November 2022].

\bibitem{NISTC3}
NI.
\newblock Ni-stc3 timing and synchronization technology.
\newblock
  \url{https://www.ni.com/en-us/shop/pxi/ni-pxi-modular-instrument-design-advantages.html},
  2022.

\bibitem{zeroMQ}
ZeroMQ.
\newblock 10 gbe tests with zeromq v4.3.2.
\newblock \url{http://wiki.zeromq.org/results:10gbe-tests-v432}.
\newblock [Online; accessed 12th November 2022].

\bibitem{astor}
Tango~Controls Collaboration.
\newblock Astor (tango manager).
\newblock
  \url{https://tango-controls.readthedocs.io/en/latest/tools-and-extensions/built-in/astor/index.html}.
\newblock [Online; accessed 22nd April 2023].

\bibitem{jive}
Tango~Controls Collaboration.
\newblock Jive.
\newblock
  \url{https://tango-controls.readthedocs.io/en/latest/tools-and-extensions/built-in/jive/index.html}.
\newblock [Online; accessed 22nd April 2023].

\bibitem{atkpanel}
Tango~Controls Collaboration.
\newblock Atkpanel: Tango controls official documentation.
\newblock
  \url{https://tango-controls.readthedocs.io/en/latest/tools-and-extensions/built-in/atkpanel/atkpanel.html}.
\newblock [Online; accessed 22nd April 2023].

\bibitem{taurus}
Taurus Team.
\newblock Taurus home page.
\newblock \url{https://www.taurus-scada.org}.
\newblock [Online; accessed 22nd April 2023].

\bibitem{tangoTaranta}
Taranta~collaboration Max-IV Instite~SKA.
\newblock Taranta (webjive until v 1.1.5).
\newblock \url{https://webjive.readthedocs.io/en/latest/what_is_it.html}, 2023.

\bibitem{Waltz}
Waltz.
\newblock Waltz- a platform for tango controls web application.
\newblock \url{https://waltz-docs.readthedocs.io/en/latest}.
\newblock [Online; accessed 12th November 2022].

\bibitem{Ada}
Adacore.
\newblock Adacore web server.
\newblock \url{https://docs.adacore.com/aws-docs/aws}.
\newblock [Online; accessed 12th November 2022].

\bibitem{plotly}
plotly.
\newblock plotly.
\newblock \url{https://plotly.com/python}.
\newblock [Online; accessed 12th November 2022].

\bibitem{APIRest}
Tango~Controls Collaboration.
\newblock Api rest.
\newblock
  \url{https://tango-controls.readthedocs.io/en/latest/development/advanced/rest-api.html}.
\newblock [Online; accessed 12th November 2022].

\bibitem{Trebino1997}
Rick Trebino, Kenneth~W. DeLong, David~N. Fittinghoff, John~N. Sweetser,
  Marco~A. Krumbügel, Bruce~A. Richman, and Daniel~J. Kane.
\newblock {Measuring ultrashort laser pulses in the time-frequency domain using
  frequency-resolved optical gating}.
\newblock {\em Review of Scientific Instruments}, 68(9):3277--3295, 09 1997.

\bibitem{TangoMailingList}
Tango~Controls Collaboration.
\newblock Tango controls archived mailing list.
\newblock \url{https://lists.tango-controls.org/wws/arc/info}.
\newblock [Online; accessed 22nd April 2023].

\bibitem{TangoSlack}
Tango~Controls Collaboration.
\newblock Tango controls slack channel.
\newblock \url{https://www.tango-controls.org/community/slack}.
\newblock [Online; accessed 22nd April 2023].

\bibitem{tangoinstall}
Tango~Controls Collaboration.
\newblock Tango controls official documentation: Installation.
\newblock
  \url{https://tango-controls.readthedocs.io/en/latest/installation/index.html}.
\newblock [Online; accessed 13th January 2023].

\bibitem{controlworkshop2022}
Lpa online workshop on control systems and machine learning (jan 24 – 27,
  2022).
\newblock \url{https://indico.physik.uni-muenchen.de/event/135}.
\newblock [Online; accessed 13th January 2023].

\bibitem{Wilkinson2016}
Mark~D. Wilkinson, Michel Dumontier, {IJsbrand}~Jan Aalbersberg, Gabrielle
  Appleton, Myles Axton, Arie Baak, Niklas Blomberg, Jan-Willem Boiten,
  Luiz~Bonino da~Silva~Santos, Philip~E. Bourne, Jildau Bouwman, Anthony~J.
  Brookes, Tim Clark, Mercè Crosas, Ingrid Dillo, Olivier Dumon, Scott
  Edmunds, Chris~T. Evelo, Richard Finkers, Alejandra Gonzalez-Beltran,
  Alasdair J.~G. Gray, Paul Groth, Carole Goble, Jeffrey~S. Grethe, Jaap
  Heringa, Peter A.~C. ’t Hoen, Rob Hooft, Tobias Kuhn, Ruben Kok, Joost Kok,
  Scott~J. Lusher, Maryann~E. Martone, Albert Mons, Abel~L. Packer, Bengt
  Persson, Philippe Rocca-Serra, Marco Roos, Rene van Schaik, Susanna-Assunta
  Sansone, Erik Schultes, Thierry Sengstag, Ted Slater, George Strawn,
  Morris~A. Swertz, Mark Thompson, Johan van~der Lei, Erik van Mulligen, Jan
  Velterop, Andra Waagmeester, Peter Wittenburg, Katherine Wolstencroft, Jun
  Zhao, and Barend Mons.
\newblock The {FAIR} guiding principles for scientific data management and
  stewardship.
\newblock {\em Scientific Data}, 3(1):160018, 2016.
\newblock Number: 1 Publisher: Nature Publishing Group.

\bibitem{gofair}
GO~FAIR.
\newblock Go fair official website.
\newblock \url{https://www.go-fair.org}.
\newblock [Online; accessed 13th January 2023].

\bibitem{webontlang}
W3C OWL~Working Group.
\newblock Owl 2 web ontology language.
\newblock \url{https://www.w3.org/TR/owl2-overview}.
\newblock [Online; accessed 21st December 2022].

\bibitem{Molinaro2021}
Marco Molinaro, Mark Allen, François Bonnarel, Françoise Genova, Markus
  Demleitner, Kay Graf, Dave Morris, Enrique Solano, and André Schaaff.
\newblock Supporting {FAIR} principles in the astrophysics community: the
  european experience.

\bibitem{Allen2021}
Mark~G. Allen, Stefania Amodeo, François Bonnarel, Marco Molinaro, Françoise
  Genova, Martino Romaniello, Hendrik Heinl, André Schaaff, and Soizick
  Lesteven.
\newblock Data stewardship of next-generation data products and services
  related to {ESFRI} and other astrophysics infrastructures.

\bibitem{Weeks2022}
Allen Weeks, Gábor Szabó, Roman Hvězda, Florian Gliksohn, and Teodor
  Ivănoaica.
\newblock {ELI} {ERIC} data policy.
\newblock 202.
\newblock Publisher: Zenodo.

\bibitem{fairsfair}
Implementation \& adoption stories {\textbar} {FAIRsFAIR}.

\bibitem{forbesfair}
Cloudera Contributor.
\newblock Forbes insights: The case for {FAIR} data.
\newblock Section: Innovation.

\bibitem{sheffieldfair}
{FAIR} case studies: Good practice on {FAIR} data and software.

\bibitem{openpmdwebsite}
openPMD.
\newblock openpmd github page.
\newblock \url{https://github.com/openPMD}.
\newblock [Online; accessed 13th January 2023].

\bibitem{openpmdcodes}
openPMD.
\newblock Projects using openpmd.
\newblock \url{https://github.com/openPMD/openPMD-projects}.
\newblock [Online; accessed 21st December 2022].

\bibitem{PICMI}
PICMI.
\newblock Picmi standard.
\newblock \url{https://picmi-standard.github.io}.
\newblock [Online; accessed 12th November 2022].

\bibitem{openpmdccd}
openPMD.
\newblock openpmd-ccd github page.
\newblock \url{https://github.com/openPMD/openPMD-CCD}.
\newblock [Online; accessed 13th January 2023].

\bibitem{Konnecke2015}
M.~Könnecke, F.~A. Akeroyd, H.~J. Bernstein, A.~S. Brewster, S.~I. Campbell,
  B.~Clausen, S.~Cottrell, J.~U. Hoffmann, P.~R. Jemian, D.~Männicke,
  R.~Osborn, P.~F. Peterson, T.~Richter, J.~Suzuki, B.~Watts, E.~Wintersberger,
  and J.~Wuttke.
\newblock The {NeXus} data format.
\newblock {\em Journal of Applied Crystallography}, 48(1):301--305, 2015.
\newblock Number: 1 Publisher: International Union of Crystallography.

\bibitem{Streeter2023HPLSE}
M.J.V. Streeter, C.~Colgan, C.C. Cobo, C.~Arran, E.E. Los, R.~Watt,
  N.~Bourgeois, L.~Calvin, J.~Carderelli, N.~Cavanagh, and et~al.
\newblock Laser wakefield accelerator modelling with variational neural
  networks.
\newblock {\em High Power Laser Science and Engineering}, page 1–9, 2023.

\bibitem{mlflow}
MLflow.
\newblock Mlflow.
\newblock \url{https://mlflow.org}.
\newblock [Online; accessed 12th November 2022].

\bibitem{joglekar_2022}
Archis~S. Joglekar and Alexander~G.R. Thomas.
\newblock Unsupervised discovery of nonlinear plasma physics using
  differentiable kinetic simulations.
\newblock {\em Journal of Plasma Physics}, 88(6):905880608, December 2022.

\end{thebibliography}

\end{document}